\documentclass[structabstract]{aa}
\usepackage{natbib}
\usepackage{hyperref}
\usepackage{amsmath}
\hypersetup{colorlinks=true,citecolor=blue}
\usepackage{graphicx}
\usepackage{aas_macros,color}

\newcommand{\be}{\begin{equation}}
\newcommand{\ee}{\end{equation}}
\newcommand{\ba}{\begin{align}}
\newcommand{\ea}{\end{align}}
\newcommand{\bea}{\begin{eqnarray}}
\newcommand{\eea}{\end{eqnarray}}
\newcommand{\rd}{\rm{d}}

\hypersetup{
 linkcolor=red,
 }

\begin{document}

\title{Removal of two large-scale cosmic microwave background anomalies after subtraction of the integrated Sachs-Wolfe effect}
\titlerunning{Removal of two large-scale CMB anomalies after subtraction of the ISW effect}
\author{A. Rassat \inst{1,2} \thanks{anais.rassat@epfl.ch} \and J.-L. Starck \inst{2}  \and F.-X. Dup\'e \inst{3}}
\institute{$^1$ Laboratoire d'Astrophysique, Ecole Polytechnique F\'ed\'erale de Lausanne (EPFL), Observatoire de Sauverny, CH-1290, Versoix, Switzerland.\\
$^2$ Laboratoire AIM, UMR CEA-CNRS-Paris 7, Irfu, Service d'Astrophysique, CEA Saclay, F-91191 GIF-SUR-YVETTE CEDEX, France.\\ $^3$ LIF - Qarma, UMR CNRS 7279 Aix-Marseille Univ }

\abstract
{Although there is currently a debate over the significance of the claimed large-scale anomalies in the Cosmic Microwave Background (CMB), their existence is not totally dismissed. In parallel to the debate over their statistical significance, recent work has also focussed on masks and secondary anisotropies as potential sources of these anomalies.}
{In this work we investigate simultaneously the impact of the method used to account for masked regions as well as the impact of the integrated Sachs-Wolfe (ISW) effect, which is the large-scale secondary anisotropy most likely to affect the CMB anomalies. In this sense, our work is an update of both Francis \& Peacock 2010 and Kim et al. 2012. Our aim is to identify trends in CMB data from different years and with different mask treatments.}
{We reconstruct the ISW signal due to 2 Micron All-Sky Survey (2MASS) and NRAO VLA Sky Survey (NVSS) galaxies, effectively reconstructing the low-redshift ISW signal out to $z\sim1$. We account for regions of missing data using the sparse inpainting technique of Abrial et al. 2008, Starck, Murtagh \& Fadili 2010 and Starck, Fadili \& Rassat 2012. We test sparse inpainting of the CMB, Large Scale Structure and ISW and find that it constitutes a bias-free reconstruction method suitable to study large-scale statistical isotropy and the ISW effect.}
{We focus on three large-scale CMB anomalies: the low quadrupole, the quadrupole/octopole alignment, and the octopole planarity.  After sparse inpainting, the low quadrupole becomes more anomalous, whilst the quadrupole/octopole alignment becomes less anomalous. The significance of the low quadrupole is unchanged after subtraction of the ISW effect, while the trend amongst the CMB maps is that the quadrupole/octopole alignment has reduced significance, yet other hypotheses remain possible as well (e.g. exotic physics). Our results also suggest that both of these anomalies may be due to the quadrupole alone. The octopole planarity significance is also reduced after inpainting and after ISW subtraction, however, we do not find that it was very anomalous to start with. In the spirit of participating in reproducible research, we make all codes and resulting products which constitute main results of this paper public here: \url{http://www.cosmostat.org/anomaliesCMB.html}. }
{}

\maketitle
\section{Introduction}

In recent years, cosmological observations \citep{WMAP7,Percival:2007b,Schrabback:2010} have lead to the establishment of a standard cosmological model.
This model assumes an inflationary scenario leading to Gaussian features in the temperature anisotropies of the Cosmic Microwave Background (CMB). Since the advent of the COsmic Background Explorer \citep[COBE,][]{COBE:1990} and Wilkinson Microwave Anisotropy Probe \citep[WMAP,][]{Spergel:2003cb} several signatures of lack of statistical isotropy, or ``anomalies'', have been reported on large scales.

On the largest scales, a low quadrupole was reported with COBE data \citep{Hinshaw:1996,Bond:1998} and later confirmed with WMAP data \citep{Spergel:2003cb}, suggesting that it was not due to Galactic emissions. The octopole presented an unusual planarity and a correlation with the quadrupole \citep{Tegmark:2003ve,OctPlanarity,Slosar:2004s,Copi:2010}. Other anomalies include a north/south power asymmetry \citep{Eriksen:2003db,Bernui:2006}, an anomalous ``cold spot'' in the CMB \citep{Vielva:2004,Cruz:2005,Cruz:2006sv}, alignments of other large-scale multipoles \citep{Schwarz:2004,Copi:2005ff}, the so-called Axis of Evil \citep{Land:2005ad}, and other violations of statistical isotropy \citep{Hajian:2003s,Land:2004bs}. 

If confirmed, these anomalies could provide a new window into exotic early-Universe physics. However, there is much debate about the possible causes of these claimed anomalies. The statistics used to detect them are often subtle, and given the large cosmic variance on the scales considered, the anomalies could be due to a simple statistical fluke or in fact not be anomalous \citep{WmapAnomalies,Efstathiou:2010,ilc:w7,ilc:w9}, depending on how the significance is measured. 

The anomalies could also have a low-redshift cosmological origin. Since statistical isotropy is predicted for the early Universe, analyses should focus on the primordial CMB, i.e. one from which secondary low-redshift cosmological signals have been removed. This should be done \emph{whether or not} one believes the significance of the reported anomalies in the CMB. 

 \cite{Hiranya:ksz} investigated the kinetic Sunyaev-Zeldovich (kSZ) effect, a known secondary anisotropy, and found it was unlikely that the kSZ effect was the origin of these anomalies. On the largest scales, the only secondary anisotropy is the integrated Sachs-Wolfe (ISW) effect, which is correlated with foreground large-scale structure.  A first study by \cite{Rudnick:2007} detected a cold spot in the NRAO VLA Sky Survey (NVSS), similar to that found in WMAP data, suggesting that the WMAP cold spot could be due to the late ISW effect. However, \cite{Smith:2010Huterer} showed that the cold spot in the NVSS was no longer significant when systematics were taken into account, and \cite{Granett:2009,Granett:2010} rule out that the cold spot be due to a supervoid. Recently, \cite{CMBSN:2012} found a correlation between the spatial distribution of supernovae and the CMB, which they claim could be due to dust contribution, ionised gas, or the ISW effect.  

\cite{Francis:2010iswanomalies} (hereafter FP10) investigated the impact of the ISW field due to 2 Micron All-Sky Survey (2MASS) galaxies on several CMB anomalies: the low quadrupole, the quadrupole/octopole alignment, the planarity of the octopole, the north/south asymmetry and the cold spot. They found that the first two of these anomalies were reduced in significance after the removal of the foreground signal, but that the octopole planarity was actually increased. The north/south asymmetry was somewhat reduced and the cold spot remained anomalous.

The study of these large-scale anomalies
is extremely sensitive to the treatment of the Galactic mask to account for foreground removals \citep{Bernui:2006,Slosar:2004s,ilc:w9} in both the CMB and galaxy survey data used to reconstruct the ISW field, hence a statistically un-biased method for dealing with missing data should be used. Regarding the CMB, \cite{ilc:w9} found that the limiting factor in measuring the quadrupole/octopole alignment was foreground removal. Regarding the LSS side, FP10 reconstructed missing data by first filling the masked region with a random Poisson sampling of galaxies with the same average number density as outside the mask and then applying a Wiener filter, which is optimal for Gaussian data. \cite{Kim:2012} investigated methods to reconstruct CMB data in masked regions using a Gaussian-constrained harmonic inpainting method. They found that the quadrupole/octopole alignment in CMB maps was increased after inpainting treatment. Both methods assume Gaussianity of the underlying maps, which is a limitation in the context of search for statistical anisotropy. The filling of the masked region with random galaxies may also cause an artificial quadrupole since the Galactic mask has the shape of a quadrupole.

In this paper,  we investigate the use of sparse inpainting (see \cite{Abrial2008} and \cite*{starck:book10}, hereafter A08-SMF10) on CMB (WMAP) and LSS maps (2MASS and NVSS) in the context of probing large-scale anomalies in the CMB. The sparse inpainting does not assume that the true underlying map is Gaussian, or statistically isotropic. We reconstruct a full-sky tomographic ISW map due to 2MASS and NVSS galaxies. We test whether removing the ISW signal due to these LSS maps affects the observed large-scale anomalies in the CMB by focussing on the same three large-scale anomalies as FP10.

This paper updates the work presented in \cite{Francis:2010iswanomalies} and \cite{Kim:2012}, with the following differences:
\begin{enumerate}
\item \emph{Tomography}: We reconstruct the ISW signal due to both 2MASS \citep{Jarrett:2000me} and NVSS \citep{NVSS} data. The data sets are described in Section \ref{sec:data}
\item \emph{Trends}: We look for trends by considering 11 different WMAP data sets, described in Section \ref{sec:data}

\item \emph{Sparse inpainting}: Missing data in CMB and LSS maps are reconstructed using the sparse inpainting technique of A08-SMF10 and \cite*{Starck:2012}, described in Section \ref{sec:inpainting}, which does not assume that the data is Gaussian or statistically isotropic

\item \emph{Tests for biases}: We test that the sparse inpainting method does not introduce biases in statistical isotropy tests nor creates a spurious ISW signal in Section \ref{sec:testinpainting}
\end{enumerate}

In Section \ref{sec:reconstructing} we present how it is possible to estimate the primordial CMB on large scales with knowledge of foreground LSS maps by reconstructing the ISW field.  In Section \ref{sec:probing}, we investigate the effect of subtracting the ISW field due to 2MASS and NVSS galaxies on three previously reported large-scale anomalies: the low quadrupole, the quadrupole/octopole alignment and the octopole planarity. In Section \ref{sec:discussion}, we present a discussion of our results.

\section{Estimating the Large-Scale Primordial CMB}\label{sec:reconstructing}

\begin{table*}[htbp]
   \centering
   \begin{tabular}{@{} lcll @{}} 
   \hline
Label & WMAP Year &Mask Treatment & Reference\\
\hline
TOH W1 & W1&ILC&\cite{Tegmark:2003ve}\\
ILC W3 & W3 &ILC&\cite{ilc:w3}\\
ILC W5 & W5&ILC&\cite{ilc:w5}\\
ILC W7 &W7 &ILC&\cite{ilc:w7}\\
ILC W9 & W9 & ILC&\cite{ilc:w9}\\
TOH W1 (inp) & W1&ILC + Sparse Inpainting&This work\\
ILC W3 (inp) &W3 & ILC + Sparse Inpainting&This work\\
ILC W5 (inp) &W5 & ILC + Sparse Inpainting&This work\\
Dela W5 (inp) &W5&ILC-Wavelets + Sparse Inpainting &This work\\
ILC W7 (inp) &W7& ILC + Sparse Inpainting&This work\\
ILC W9 (inp) &W9& ILC + Sparse Inpainting&This work\\
\hline
  \end{tabular}
  \caption{: List of the 11 temperature anisotropy maps we use to probe the anomalies in the primoridal CMB, including year of WMAP data they correspond to and a description of the mask treatment. `TOH' corresponds to the treatment in \cite{Tegmark:2003ve},`ILC' to the Internal Linear Combination method \cite[see][]{ilc:w3,ilc:w9,ilc:w5,ilc:w7}, `ILC-Wavelets' corresponds to the method in \cite{Delabrouille:2008}, and the sparse inpainting method is described in A08-SMF10 and uses the sparsity prior described in \cite*{Starck:2012}.}
  \label{tab:cmbdata}
\end{table*}

\subsection{The large-scale primordial CMB after subtraction of the integrated Sachs-Wolfe effect} 
The observed temperature anisotropies in the CMB, $\delta_{\rm OBS}$, can be described as the sum of several components: 
\begin{equation}
  \label{eq:7}
  \delta_{\mathrm{OBS}} = \delta_{\rm prim} +  \delta^{\rm total}_{\mathrm{ISW}}+\delta_{\mathrm{other}}+\mathcal{N}~,
\end{equation} where $\delta_{\rm prim}$ are the primordial temperature anisotropies, $\delta^{\rm total}_{\rm ISW}$ are the total secondary temperature anisotropies due to the ISW effect, $\delta_{\rm other}$ are other secondary anisotropies (e.g., Sunyaev-Zeldovich effect), and $\mathcal{N}$ is noise. Here, we assume that the observed data are clean of any foregrounds and that any regions requiring masking due to contaminated data will be corrected for during the sparse inpainting phase. On large scales, the ISW signal is the only secondary anisotropy and the anisotropies are cosmic variance limited, so that the last two terms can be ignored:

\begin{equation}
  \delta_{\mathrm{OBS}} \simeq \delta_{\rm prim} +  \delta^{\rm total}_{\mathrm{ISW}},{\rm ~for~large~scales.}
\end{equation}
The ISW effect arises in universes where the cosmic potential decays at late times, as is the case with dark energy, open curvature, or possibly some modified gravities. The temperature anisotropies due to the ISW effect are given by: 
\begin{equation} \delta^{\rm total}_{\rm ISW}=-2\int_{\eta_L}^{\eta_0}\Phi'\left((\eta_0-\eta)\hat{n},\eta \right)\rd\eta, \label{sec:theory:eq:isw} \end{equation} where $\eta$ is the conformal time, defined by $\rd \eta = \frac{\rd t}{a(t)}$, and $\eta_0$
and $\eta_L$ represent the conformal times today and at the surface of last scattering respectively.  The unit vector $\hat{\bf n}$ is along the line of sight and the gravitational potential
$\Phi({\bf x}, \eta)$ depends on position and time.  The integral depends on the rate of change of the potential $\Phi'=\rd \Phi / \rd\eta$. The potential field can be related to the matter field, of which a galaxy map is assumed to be a tracer.

By subtracting the reconstructed ISW signal, $\hat{\delta}_{\rm ISW}$, due to the matter field traced by a foreground galaxy survey, we can estimate the large-scale primordial CMB temperature anisotropy field $\hat{\delta}_{\rm prim}$ by: 
\begin{equation}\hat{\delta}_{\rm prim} \simeq \delta_{\rm OBS} - \hat{\delta}_{\rm ISW} {\rm ,~for~large~scales},\end{equation} where this relation tends to equality (on large scales) when the entire mass distribution of the Universe is used to reconstruct the ISW field, i.e. when $ \hat{\delta}_{\rm ISW} \equiv  \delta^{\rm total}_{\rm ISW}$.

In this paper, we estimate $\hat{\delta}_{\rm prim}$ by subtracting the ISW temperature anisotropy field as reconstructed in Section \ref{sec:inpainting} using 2MASS and NVSS data (see Section \ref{sec:data}):
\begin{equation} \hat{\delta}_{\rm prim}\simeq \delta_{\rm OBS}-\hat{\delta}_{\rm ISW}^{\rm 2MASS}-\hat{\delta}_{\rm ISW}^{\rm NVSS}-\delta_{kD,\ell=2}, \label{eq:dt1}\end{equation} where the extra last term corresponds to the removal of the kinetic Doppler quadrupole (ignoring the monopole and dipole). The kinetic Doppler quadrupole contribution is given by: 
\begin{equation} \delta_{kD,\ell=2}=\left(\frac{v}{c}\right)^2\left[cos^2 \theta-\frac{1}{3}\right],\end{equation} where $\theta$ is the angle between the position on the sky and the direction of motion creating the kinetic Doppler quadrupole \citep{Copi:2005ff}. For our calculations, we take $v=370 {\rm km} s^{-1}$ towards $(l,b)=(263.85^\circ,48.25^\circ$), and $c$ is the speed of light, similarly to what is done in \cite{Francis:2010iswanomalies}.  We make the kinetic Doppler quadrupole map in Healpix format as well as code to generate it available for download here \url{http://www.cosmostat.org/anomaliesCMB.html}.

\subsection{The ISW temperature field from LSS maps}\label{sec:theory}

The temperature ISW field can be reconstructed in spherical harmonics, $\delta_{\ell m}^{\rm ISW}$, from the LSS field $g_{\ell m}$ \citep{Boughn:1998,Cabre:2007,Giannantonio:2008}: 
\begin{equation}
  \delta^{\rm ISW}_{\ell m} = \frac{C_{gT}(\ell)}{C_{gg}(\ell)} g_{\ell m},\label{eq:alm_isw}
\end{equation}
where $g_{\ell m}$ represent the spherical harmonic coefficients of a galaxy overdensity field $g(\theta, \phi)$, given by
\begin{equation} g(\theta, \phi) = \sum _{\ell, m} g_{\ell m} Y_{\ell m}(\theta, \phi),\end{equation} where $Y_{\ell m}(\theta,\phi)$ are the spherical harmonics. 
The spectra $C_{gg}$ and $C_{gT}$ are the galaxy (g) and CMB (T) auto- and cross-correlations measured from the data or their theoretical values given by: \begin{equation}C_{gT}(\ell) = 4 \pi b_g\int \rd k \frac{\Delta^2(k)}{k}  W_g(k)W_T(k),\label{eq:cgt}\end{equation}
\begin{equation}C_{gg}(\ell) = 4 \pi b^2_g\int \rd k \frac{\Delta^2(k)}{k}  \left[W_g(k)\right]^2, \label{eq:cgg}\end{equation}
where 
\begin{equation}W_g(k) =\int \rd r \Theta(r) j_\ell(kr) D(z),\end{equation}
\begin{equation}W_T(k) = -\frac{3\Omega_{m,0} H_0^2}{k^2c^3}\int_0^{z_L} \rd r j_\ell(kr)H(z)D(z)(f-1),\end{equation}
\begin{equation}\Delta^2(k) = \frac{4\pi}{(2\pi)^3}k^3P(k),\end{equation}
\begin{equation}\Theta(r) = \frac{r^2n(r)}{\int \rd r r^2 n(r)},\end{equation} where we use the same notation as in \cite{Rassat:2007KRL} and have assumed a linear bias $b(k,z)\equiv b_g$ and $D(z)$ is the linear growth function.

From Equation \ref{eq:alm_isw}, we can check that: 
\begin{equation}\left<\delta^{\rm ISW}_{\ell m}g^*_{\ell m}\right>=C_{gT}(\ell), \end{equation}
and 
\begin{eqnarray}\left<\delta^{\rm ISW}_{\ell m}\delta^{*{\rm ISW}}_{\ell m}\right>=  \frac{C^2_{\rm gT}(\ell)}{C_{\rm gg}(\ell)}=C_{\rm ISW}(\ell) .\label{eq:iswspectrum}\end{eqnarray}
Equation \ref{eq:alm_isw} and \ref{eq:iswspectrum} show that the ISW temperature field is independent of the galaxy bias $b_g$, which relates the galaxy and matter (m) fluctuations by $g_{\ell m} = b_g\delta_{{\rm m},\ell m}$, and that it is not necessary to estimate the value of the galaxy bias in order to estimate the ISW temperature field.  

We note that the above method is suitable for photometric surveys where a projected spherical harmonic decomposition is satisfactory. For spectroscopic surveys, a full three-dimensional reconstruction can be done in spherical Fourier-Bessel decomposition, as in \cite{ISW3D} and \cite{Rassat:2012bao} for the theoretical calculations, and using a code such as 3DEX \citep{3DEX} for the measurements.

\begin{figure*}[htbp]
   \centering
         \includegraphics[width=8cm]{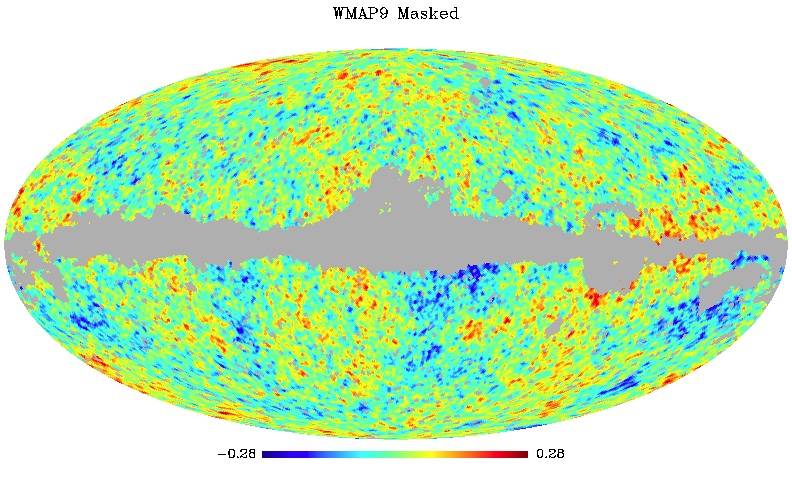}\includegraphics[width=8cm]{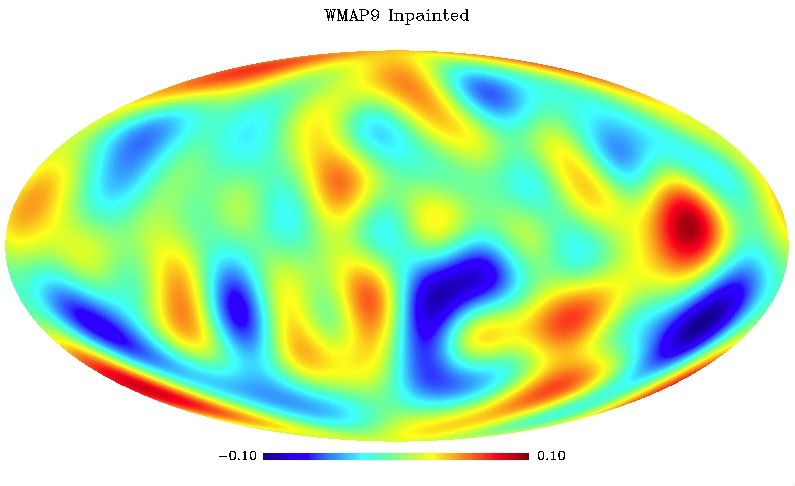}
      \includegraphics[width=8cm]{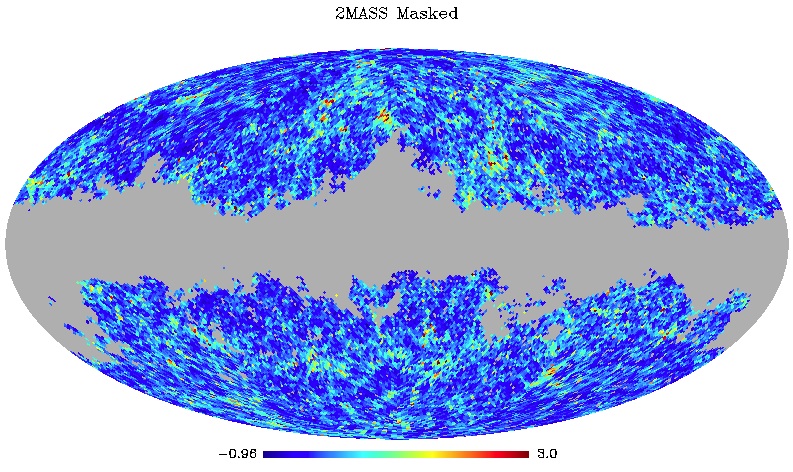}\includegraphics[width=8cm]{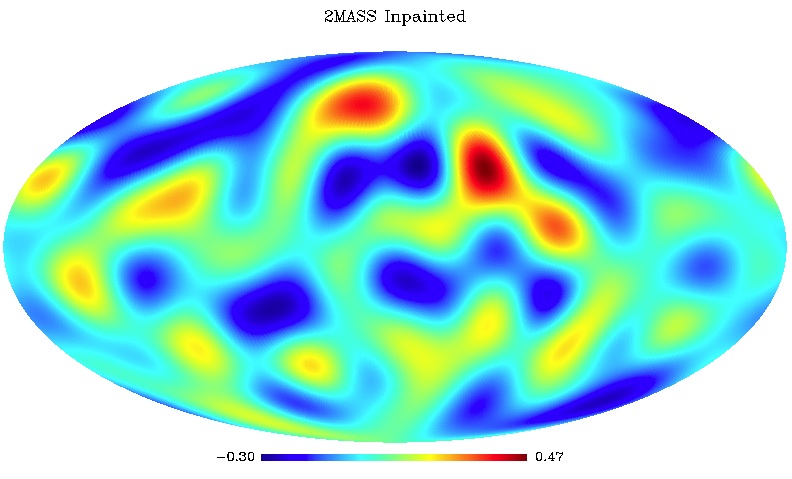}
            \includegraphics[width=8cm]{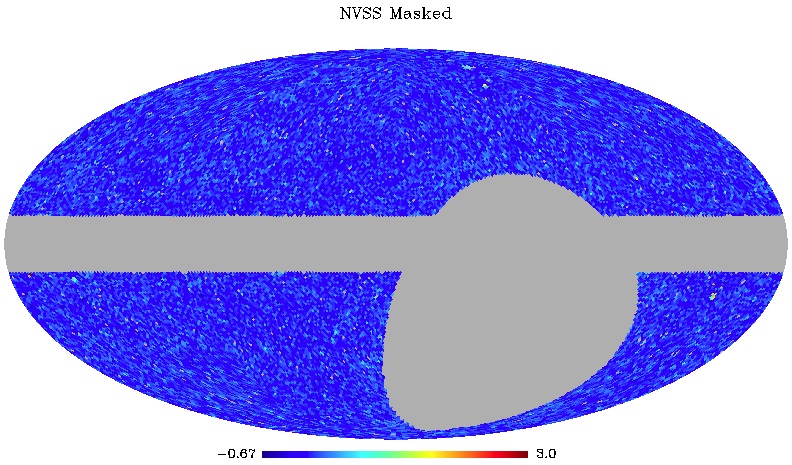}\includegraphics[width=8cm]{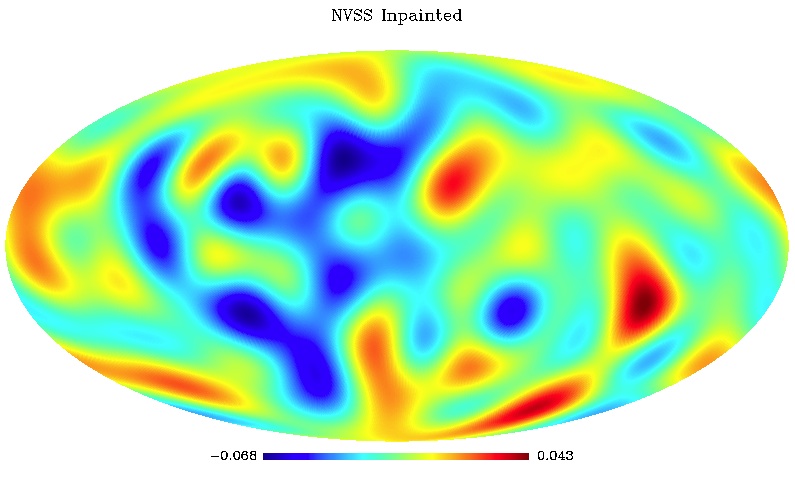}
   \caption{ WMAP ILC 9th year CMB data (top in $mK$), 2MASS data (middle), and NVSS data (bottom) with mask (left) and after sparse inpainting (right). Data in the right column is presented up to $\ell=10$ for all maps. The original 2MASS data (middle left) is plotted with a maximum overdensity value of 3 to increase contrast in the map.}
   \label{fig:maps}
\end{figure*}

\section{Data}\label{sec:data}
We are interested in reconstructing the temperature ISW fields due to 2MASS and NVSS data. We describe the CMB and LSS data we use in the following section, while the reconstruction is described in Section \ref{sec:inpainting}.

\subsection{Cosmic Microwave Background Data}\label{sec:wmapdata}
For the CMB, we investigate data from several years of WMAP, including independent treatments of masked data. We do this to identify trends in the observed large-scale anomalies. We use 11 maps in total: : the \cite{Tegmark:2003ve} reduced-foreground CMB map (TOH); the Internal Linear Combination Map (ILC) WMAP data from the 3rd year  \citep{ilc:w3} \footnote{\url{http://lambda.gsfc.nasa.gov/product/map/dr2/ilc_map_get.cfm}}, 5th year \citep{ilc:w5}\footnote{\url{http://lambda.gsfc.nasa.gov/product/map/dr3/ilc_map_get.cfm}}, 7th year \citep{ilc:w7}\footnote{\url{http://lambda.gsfc.nasa.gov/product/map/dr4/ilc_map_get.cfm}}, and 9th year \citep{ilc:w9} \footnote{\url{http://lambda.gsfc.nasa.gov/product/map/dr5/ilc_map_get.cfm}} as well as sparsely inpainted versions of the ILC and TOH maps (see Section \ref{sec:inpainting}).  We also include the sparsely inpainted WMAP 5th year ILC map by \cite{Delabrouille:2008}, which was reconstructed using a wavelet technique.  We summarise the CMB maps used in Table \ref{tab:cmbdata}. The ILC 9th year temperature overdensity map is shown with its corresponding mask \footnote{We use the 7th year temperature analysis mask for $nside=512$ (\url{http://0-lambda.gsfc.nasa.gov.iii-server.ualr.edu/product/map/dr4/masks_get.cfm})} in the left-hand side of Figure \ref{fig:maps} (top).

\subsection{The 2 Micron All Sky Survey (2MASS) Data}\label{sec:2mass}
As a tracer of the low-redshift matter distribution, we use the publicly available 2MASS full-sky extended source catalogue (XSC) selected in the near IR \citep{Jarrett:2000me}, which has median redshift $\bar{z}\sim0.07$.  The near-IR selection means galaxies are relatively well observed, even in the region of the Galactic plane. As a result, the 2MASS survey has a large sky coverage, which is ideal for studying the ISW effect, with $f^{\rm 2MASS}_{\rm sky} = 0.69$. For exact details of our selection criteria and the mask, see \cite{Rassat:2007KRL} and Section 6.2 of \cite{Dupe:2010}. The 2MASS overdensity map is shown with its corresponding mask on the left-hand side of Figure \ref{fig:maps} (middle). 

\subsection{ The NRAO VLA Sky Survey (NVSS) Data}\label{sec:nvss}
As a tracer of the intermediate redshift matter distribution, we use the publicly available NRAO VLA Sky Survey (NVSS), which has a median redshift $\bar{z}\sim1.4$. The NVSS survey is a 1.4GHz (radio) continuum survey covering the sky north of $\delta=-37^\circ$, in which most sources away from the Galactic plane are of extragalactic origin \citep*{Smith:2007}.  We select only sources with flux larger than $2.5$ mJy. For the mask we exclude the Galactic region with $|b|<10^\circ$ and also mask out a $0.6^\circ$ radius around bright sources with flux larger than $2.5$ Jy, as in \cite{Ho:2008}. The resulting mask gives us an effective sky fraction of $f^{\rm NVSS}_{\rm sky}=0.66$.

In order to account for the problem of declination-dependent density, which NVSS data suffers from, we apply the same method as in \cite{Schiavon:2012} and \cite{Vielva2006}: the NVSS overdensity is calculated separately for nine declination bins with $\Delta\sin(\delta)=0.1$.  

Alternatives to the declination problem have been proposed, e.g. by considering only objects above $5$mJy \citep{Barreiro:2012} or including templates to project out the declination-striping modes \citep{Ho:2008}. However, none of these methods satisfactorily solves a problem of missing power on large scales for NVSS \citep{Hernandez:2009}. The NVSS overdensity map is shown with its corresponding mask on the left-hand side of Figure \ref{fig:maps} (bottom).

\section{Reconstruction of Full-Sky Maps} \label{sec:inpainting}
Large-scale modes in the CMB and LSS fields are very sensitive to large coherent regions of missing data, such as those which are due to a Galactic mask. Therefore it is crucial to use a reconstruction method which does not introduce biases in the reconstructed field. To account for regions of missing data in CMB and LSS maps, we use sparse inpainting (A08-SMF10) to reconstruct regions of missing data \citep[see also Appendix A in][]{Dupe:2010}.  The sparse inpainting approach is powerful as the only assumption it makes on the underlying field is sparse representation. In this particular case, the assumption is therefore that the CMB and LSS signals are sparsely represented in spherical harmonic space, i.e. only a few $a_{\ell m}$s (not $C(\ell)$s) are required to describe the data. This is easily verified directly from the data and is described in more detail in \cite{Starck:2012} and \cite{Dupe:2010}. 

Sparse inpainting does not assume statistical isotropy or Gaussianity of the underlying field, unlike other methods such as Wiener filtering \cite[used by ][]{Francis:2010iswanomalies} or constrained Gaussian realisations \citep[as used by][]{Kim:2012}.  

In section \ref{sec:testinpainting}, we test whether the sparse inpainting reconstruction might produce biases in measurements of statistical isotropy or a spurious ISW signal. In section \ref{sec:tracers}, we describe the details of the sparse inpainting reconstruction for the LSS and CMB maps and in section \ref{sec:reconstruction:isw}, we describe how we use these to reconstruct full-sky ISW maps. 

\subsection{Sparse Inpainting as a Bias-Free Reconstruction Method}\label{sec:testinpainting}
In order to use the reconstructed maps to test large-scale anomalies in the CMB, we must first be confident that there are no spurious large-scale anomalies due to our reconstruction technique. The full details of the tests are given in Appendix \ref{app:inpainting}, and we highlight the main conclusions here.

We first test whether sparse inpainting affects the three anomalies we are testing in the CMB data (see Section \ref{sec:probing} for the anomaly descriptions). We test this by considering two sets of Gaussian random field simulations of CMB data. The first set of simulations has a low quadrupole as theoretical input, and the second set has a WMAP7 best-fit theoretical quadrupole (see Equation \ref{eq:bestfitquad}). We test for the three anomalies for both sets of simulations before and after inpainting (Tables \ref{tab:app:cmb1}, \ref{tab:app:cmb2}). We also apply the same tests to Gaussian random field simulations of 2MASS and NVSS galaxy fields, using theoretical values of their spectra (including the galaxy bias, Tables \ref{tab:app:2mass} and \ref{tab:app:nvss}).

We find that the sparse inpainting method applied to statistically isotropic CMB and LSS simulated data does not introduce any significant biases for any of the three tests of statistical isotropy. After inpainting, the quadrupole is slightly lower but only by 3\%, whether considering the simulations with low quadrupole or theoretical quadrupole on input. The test of quadrupole/octopole alignment can be somewhat altered due to sparse inpainting (see the standard deviation on the bias). We show here that if the CMB and LSS fields are statistically isotropic, then sparse inpainting does not alter this. 

Finally, we want to test whether sparse inpainting will alter in any way the reconstructed ISW temperature signal, either by creating a spurious ISW signal or biasing the tests of statistical isotropy. To do this, we first consider simulated NVSS maps with uncorrelated CMB maps (1st line of Table \ref{tab:app:nvssisw}). We test for cross-correlations before and after inpainting of the individual NVSS and CMB maps (which have different masks). Since the maps are statistically uncorrelated, we expect a null cross-correlation on average, which we find is the case before and after applying sparse inpainting. This shows that sparse inpainting does not introduce a spurious ISW or cross-correlation signal, confirming what we had found in \cite{Dupe:2010}. 

We also test the quadrupole/octopole alignment and octopole planarity of the reconstructed ISW signal from NVSS galaxies before and after inpainting (again where the NVSS and CMB maps are inpainting separately with different masks, see Table \ref{tab:app:nvssisw}) and find that the tests are unbiased after sparse inpainting.

We note that it would, of course, be interesting to know whether sparse inpainting is also unbiased in the case of statistically anisotropic fields, for example, the case where the CMB really does have anomalies before sparse inpainting is applied. However to do this, one would first have to decide on the models causing the anomalies (exotic physics, foregrounds, etc\ldots) in order to test realistic statistical anisotropy. In this paper, we focus only on finding out if sparse inpainting creates spurious statistically \emph{anisotropic} signatures in data where the true distribution is statistically \emph{isotropic}.

\subsection{Full-sky CMB and LSS maps}\label{sec:tracers}

We describe here the details for the sparse inpainting reconstruction for the CMB, 2MASS and NVSS data. Our goal is to reconstruct the quadrupole and octopole only ($\ell=2,3$), since we are only interested in anomalies on the largest scales. However, due to mode correlations induced by the masked data, we must consider multipoles larger than $\ell>3$ for the reconstruction. 

For the 11 CMB maps we reconstruct the harmonic coefficients up to $\ell=64$ using $nside=64$ using the sparsity prior as described in \cite*{Starck:2012}, i.e. using the following command line in the open-source sparse inpainting package {\tt ISAP} software {\footnotemark[8]}:\\

{\tt > alm = cmb\_lowl\_alm\_inpainting(map, Mask, lmax=lmax, niter=100, InpMap=result)}. \\

For 2MASS data, we reconstruct the harmonic coefficients up to $\ell=64$ using $nside=64$ with the following command line in {\tt ISAP} software {\footnotemark[8]}: \\

{\tt > alm = cmb\_lowl\_alm\_inpainting(t, Mask,lmax=lmax, niter=500, InpMap=result, /galaxy)},\\

where the keyword {\tt /galaxy} was optimised for point source catalogues like galaxy surveys to take into account that the field does not have a zero mean.

For NVSS data, we reconstruct the harmonics up to $\ell=64$ using $nside=128$ and the same command line and options as for 2MASS. The WMAP 9th year ILC, 2MASS and NVSS maps are shown before (left) and after inpainting (right) in Figure \ref{fig:maps}. The ten other CMB maps are also reconstructed in the same manner, but not plotted in Figure \ref{fig:maps}. 

\begin{figure}[htbp]
   \centering
   \vspace{-2.5cm}
   \hspace{-1.2cm}
   \includegraphics[width=10cm]{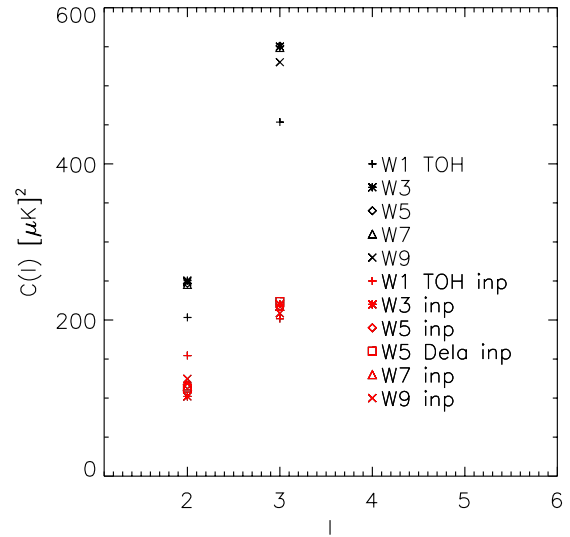} 
      \vspace{-2.5cm}
   \caption{Quadrupole and octopole values ($\mu K^2$) for the different CMB maps listed in Table \ref{tab:cmbdata}. Points in black are with no extra treatment than that given in the literature, and points in red are after application of sparse inpainting.}
   \label{fig:cmb_quad_oct}
\end{figure}

In Figure \ref{fig:cmb_quad_oct}, we plot the measured quadrupole and octopole from all CMB maps listed in Table \ref{tab:cmbdata} before (black) and after (red) sparse inpainting. Apart from the TOH W1 map, the values of the quadrupole and octopole are similar across different WMAP years. After inpainting, both the quadrupole and octopole are lowered for all maps, with the TOH W1 map again having slightly different values than the other maps for the quadrupole.

The simulations in Appendix \ref{app:inpainting} show that inpainting on W7-like maps (i.e. with low quadrupole, $C_{\ell=2}=250.6\pm161.0 ~\mu K^2$) introduces only a slight bias ($7.2$, i.e. about a 3\% drop) on the quadrupole, so that we do not expect the drop in quadrupole value shown in Figure \ref{fig:maps} to be an artifact of inpainting.

\begin{figure}[htbp]
   \centering
   \vspace{-2.5cm}
      \hspace{-1.3cm}
   \includegraphics[width=9cm]{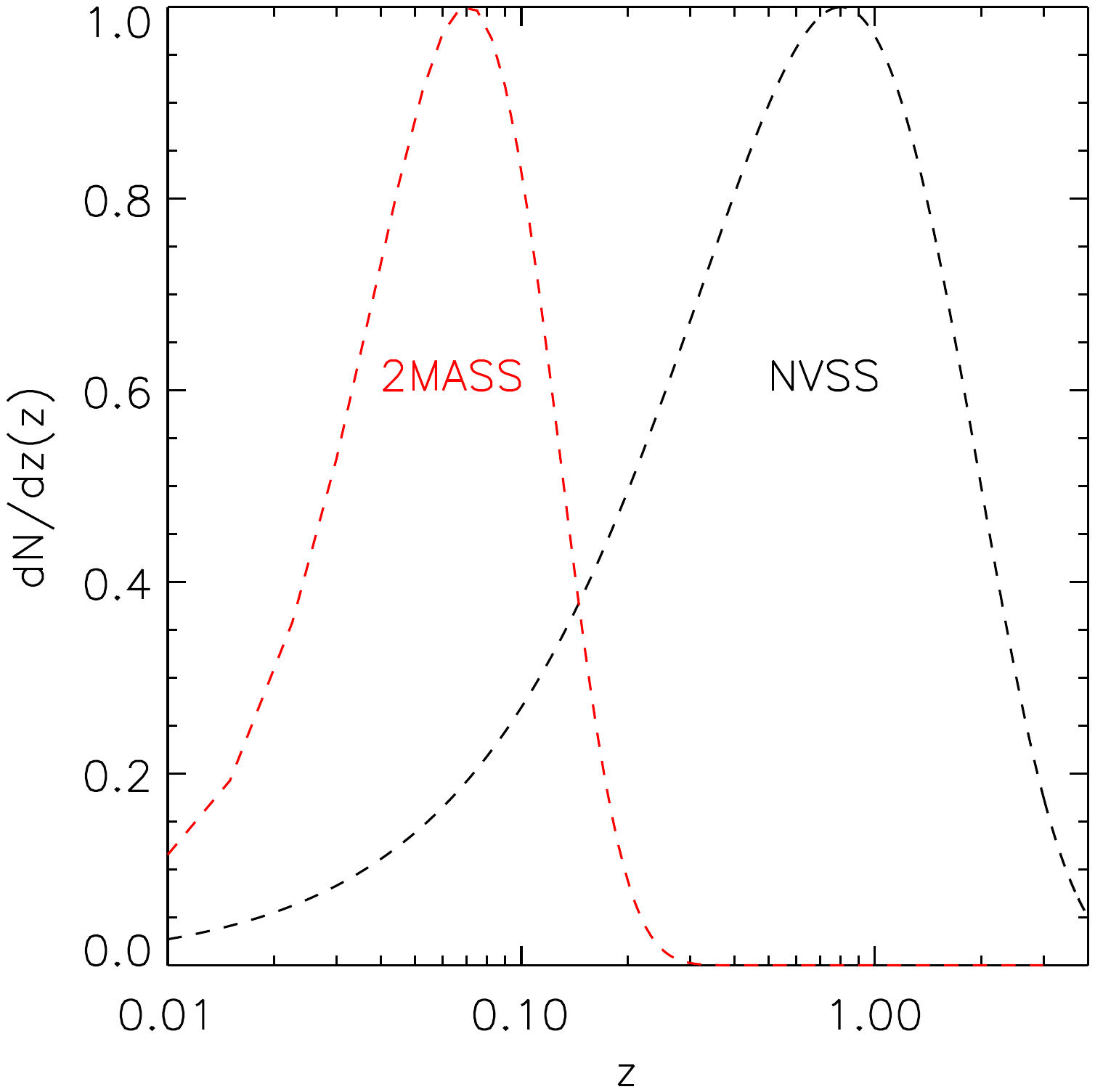} 
      \vspace{-2.5cm}
   \caption{Redshift distributions for 2MASS \cite[see][]{Afshordi:2003xu,Rassat:2007KRL,Dupe:2010} and NVSS \cite[from][]{Ho:2008}.}
   \label{fig:dndz}
\end{figure}

Regarding the LSS data, we find that after reconstruction of the NVSS map the theoretical power spectrum as predicted from the redshift distribution of radio sources given by \cite{Ho:2008} does not correspond well to the measured power spectrum, especially at higher multipoles (using their value of the galaxy bias $b_{\rm NVSS}=1.98$), as already noted by \cite{Hernandez:2009}. However, since we are focussing on very large angles only ($\ell=2,3$), it is difficult to assess whether the theoretical spectrum needs to be revisited. For this work, we use the $N(z)$ provided from \cite{Ho:2008}, which is only used in the NVSS simulations for testing sparse inpainting (Appendix \ref{app:inpainting}) and for the prediction of the value of quadrupole and octopole of the ISW due to NVSS data (Table \ref{tab:isw:th} in Section \ref{sec:reconstruction:isw}). The actual reconstructed ISW map uses the \emph{measured} auto- and cross-spectra (see the following Section \ref{sec:reconstruction:isw}) and is therefore independent of theory.

\subsection{Full-sky ISW temperature maps}\label{sec:reconstruction:isw}

We reconstruct the full-sky ISW maps due to 2MASS and NVSS data from their inpainted maps (Figure \ref{fig:maps}), using Equation \ref{eq:alm_isw}. Since the redshift distribution of 2MASS and NVSS has little overlap (see Figure \ref{fig:dndz}), we calculate the total ISW signal simply by summing the ISW maps reconstructed from each survey individually.
\begin{table}[htbp]
   \centering
   \begin{tabular}{@{} lccccc @{}} 
\hline
&$C^{\rm OBS}_{ISW}(\ell)$ ($\mu K^2$)&$C^{\rm TH}_{ISW}(\ell) (\mu K^2)$ \\
\hline
2MASS Survey\\
$\ell=2$&35.2&$12.5 \pm 9.95$\\
$\ell=3$&1.97&$3.99\pm2.52$\\

\hline
NVSS Survey \\
$\ell=2$&$5.08$&$348.5\pm 270.6$ \\
$\ell=3$&$15.6$&$127.7\pm 83.8$\\
\hline
NVSS + 2MASS\\
$\ell=2$&$41.6$&$361.0\pm 280.3$\\
$\ell=3$&$17.6$&$131.7\pm 86.4$\\
\hline

Euclid Survey\\
$\ell=2$&-&$422\pm 383$\\
$\ell=3$&-&$152\pm 138$\\
\end{tabular}
   \caption{Amplitude of the ISW temperature quadrupole and octopole due to the 2MASS galaxy survey (\emph{left}: as measured from data using inpainted WMAP 9 data for the CMB, and \emph{right}: from theoretical calculations) and a Euclid-type survey (using theoretical spectra). The standard deviations are calculated assuming $f_{\rm sky}^{\rm 2MASS}=0.69$, $f^{\rm NVSS}_{\rm sky} = 0.66$ and  $f_{\rm sky}^{\rm Euclid}=0.48$.   }
   \label{tab:isw:th}
\end{table}

\begin{figure}[htbp]
   \centering
   \vspace{-2.5cm}
   \hspace{-1.3cm}
   \includegraphics[width=10cm]{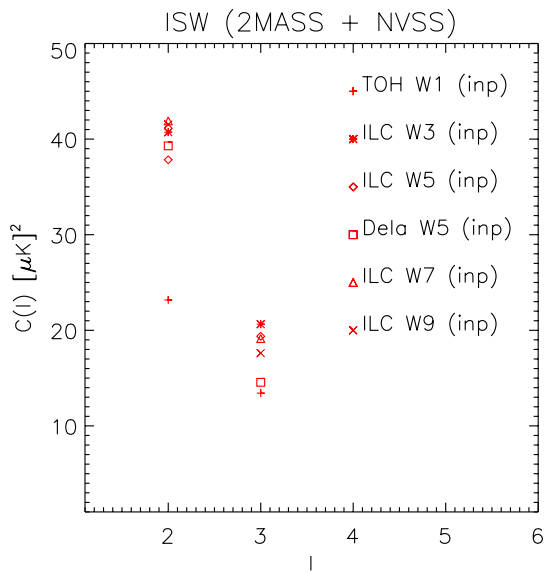} 
      \vspace{-2.5cm}
   \caption{Quadrupole and octopole values ($\mu K^2$) for the different reconstructed ISW maps. The ISW maps are reconstructed from sparsely inpainted CMB and LSS maps using Equation \ref{eq:alm_isw}, where the auto- and cross-correlations are measured directly from the data. Small variations in the phases of different renditions of CMB maps can lead to negligible differences in the auto-correlation (see Figure \ref{fig:cmb_quad_oct}) and to larger differences in a cross-correlation with a second map. This in turn can result in larger variations in the amplitude of the ISW signal, explaining why the reconstructed ISW maps vary slightly more between CMB maps than the observed temperature power does.}
   \label{fig:cmb_isw_quad_oct}
\end{figure}

We estimate the cross- and auto-correlations in Equation \ref{eq:alm_isw} using data only with inpainted maps for both the LSS and CMB data. This has the advantage of assuming only the form of the ISW signal, but not its presence, i.e. in the absence of a correlation between the CMB and the LSS maps, the reconstructed ISW map would be zero. As mentioned in Section \ref{sec:theory}, no numerical knowledge of the linear bias is necessary in this case. This method also means there is an ISW map for each CMB map considered.

Quadrupole and octopole values for the reconstructed maps using WMAP 9 ILC inpainted map are reported in Table \ref{tab:isw:th} and plotted for each CMB map in Figure \ref{fig:cmb_isw_quad_oct}.  

The theoretical values in Table \ref{tab:isw:th} are calculated employing the $N(z)$ used in \cite{Afshordi:2003xu} and \cite{Rassat:2007KRL} for 2MASS data and \cite{Ho:2008} for NVSS data. They also require the assumption of a fiducial cosmological model, for which we assume a ``vanilla" model, not a specific best-fit value, given that we are considering several CMB maps. The vanilla model we chose has the following cosmological parameters: $\Omega_m=0.25, \Omega_b=0.045, \Omega_{\rm DE}=0.75, w_0 = -1, w_a=0, n_s=1, h=0.7, \sigma_8=0.80$, i.e. a universe which can have curvature and does not have massive neutrinos. 

The theoretical values in Table \ref{tab:isw:th} are calculated using the exact formulae (not the Limber approximation). The error bars due to cosmic variance are calculated using $f_{\rm sky}^{\rm 2MASS} =0.69$, and $f_{\rm sky}^{\rm NVSS}=0.66$. 

For comparison, we also indicate the expected value of the ISW map from Euclid data \citep{Redbook,Euclidsb}. We predict this by considering a galaxy survey with median redshift $\bar{z}=0.80$ and a Smail-type redshift distribution with $\alpha=2, \beta=1.5$ \cite[see for e.g. Equation 19 in][]{DK:2012} and $f_{\rm sky}=0.48$. We calculate the expected ISW signal assuming a single large redshift bin.

Figure \ref{fig:cmb_isw_quad_oct} shows the quadrupole and octopole power for the six reconstructed ISW maps (we only consider inpainted maps for the ISW reconstruction).  The ISW quadrupole varies from $37.8-41.9$ 
$\mu K^2$ (except for TOH, which returns a significantly smaller quadrupole of $23.2$
$\mu K^2$, as shown in Figure \ref{fig:cmb_isw_quad_oct}), while the octopole varies from $13.4-20.6$
$\mu K^2$, depending on the WMAP inpainted data used. Since each map is estimated from the cross-correlation of the reconstructed LSS maps with the corresponding reconstructed CMB map, this may explain the larger variation in Figure \ref{fig:cmb_isw_quad_oct} than in Figure \ref{fig:cmb_quad_oct}. This is because a small phase change may not change the auto-correlation (as in Figure \ref{fig:cmb_quad_oct}), but can lead to larger variations in cross-correlations with another map. This in turn can lead to larger variations in the amplitude of the
 ISW signal.

In our analysis, we chose to use all six reconstructed ISW maps, because each reconstructed ISW map is self-consistently produced, i.e. without any assumption about the theoretical amplitude of the ISW signal. In addition, we are interested in identifying possible trends in the way subtraction of the ISW signal affects the CMB anomalies, which will provide a stronger case for any conclusions.

\begin{figure*}[htbp]
   \centering
         \includegraphics[width=9cm]{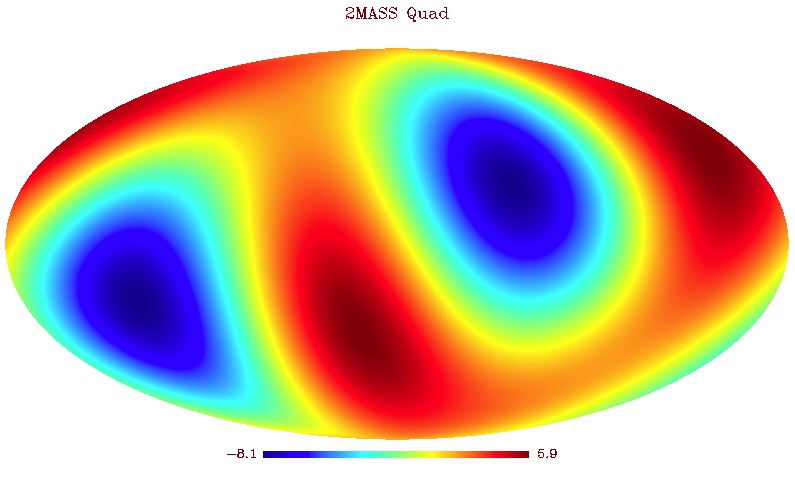}   \includegraphics[width=9cm]{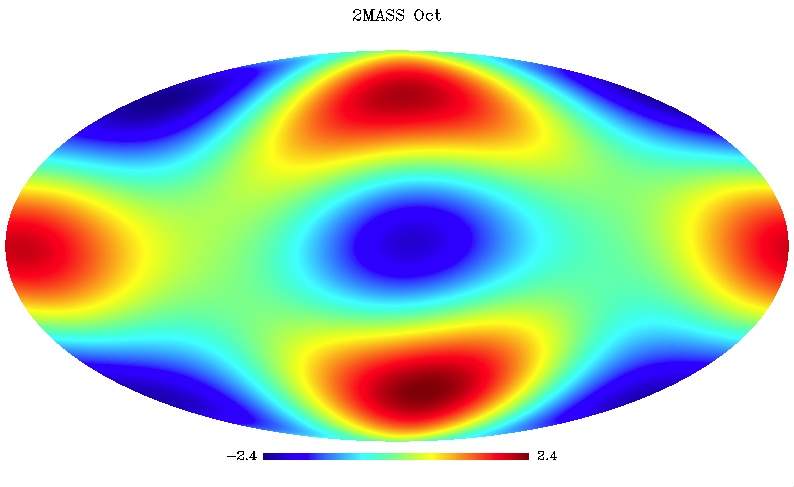}
      \includegraphics[width=9cm]{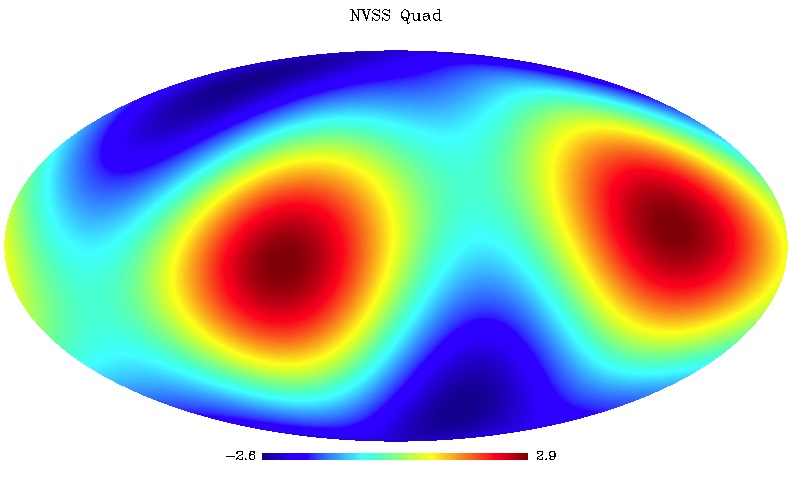}   \includegraphics[width=9cm]{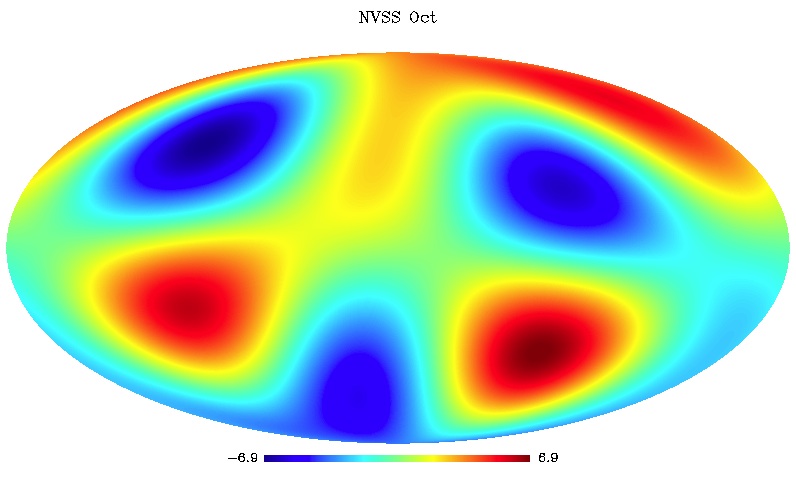}
   \includegraphics[width=9cm]{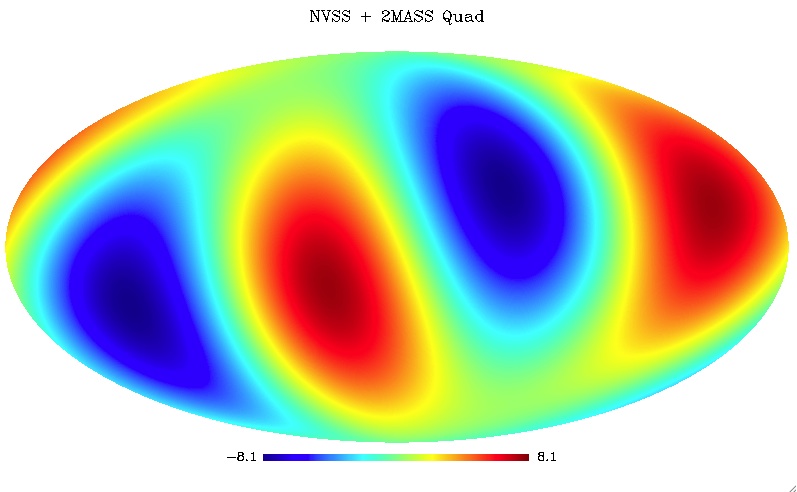}   \includegraphics[width=9cm]{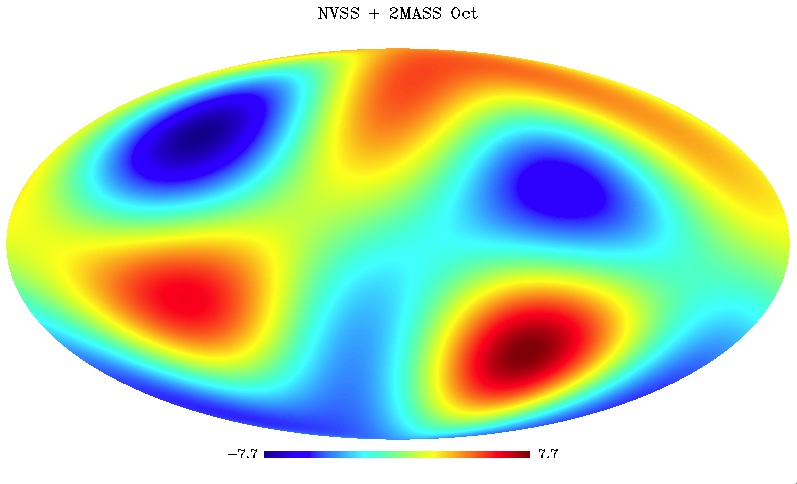}
   \caption{Quadrupole (left) and octopole (right) of the reconstructed ISW map for 2MASS (top) and NVSS (middle) and both NVSS and 2MASS together (bottom). The quadrupole and octopole of the ISW map due to 2MASS galaxies can be compared with that obtained by \cite{Francis:2010iswanomalies} (top of their Figure 3). Maps are shown in units of $\mu K$.}
   \label{fig:reconstructedmaps}
\end{figure*}

Finally, Figure \ref{fig:reconstructedmaps} shows the reconstructed ISW temperature maps due to 2MASS galaxies (top) and NVSS galaxies (middle) where the CMB map is the W7 inpainted data. Only the quadrupole (left) and octopole (right) are shown, since we are interested in probing large-scale anomalies at $\ell=2$ and $\ell=3$. 

The quadrupole and octopole of the ISW map due to 2MASS galaxies can be compared with those obtained by \cite{Francis:2010iswanomalies} (top of their Figure 3). The main difference is that our reconstructed quadrupole is much less planar. This may be due to the difference in reconstruction of missing data. \cite{Francis:2010iswanomalies} reconstructed missing data by first filling the masked region with a random Poisson sampling of galaxies with the same average number density as outside the mask. However, this can cause an artificial quadrupole, since the Galactic mask has the shape of a quadrupole.

\section{Probing Anomalies in the Observed and Primordial CMB}\label{sec:probing}
The main goal of this paper is to test whether reported large-scale anomalies in the observed CMB are still present in the primordial CMB, which we estimate by subtracting the ISW signal. Since the estimation of the ISW signal requires sparse inpainting of all CMB and LSS maps to account for regions of missing data, we can also test whether reported large-scale anomalies persist in the observed CMB after inpainting. 

We investigate three large-scale anomalies in the CMB, which are the same as those investigated in \cite{Francis:2010iswanomalies}. In Section \ref{sec:lowquad}, we consider the impact on the low quadupole power, in Section \ref{sec:alignment} the alignment of the quadrupole and octopole, and in Section \ref{sec:planarity} the planarity of the octopole. 

\subsection{Low Quadrupole Power}\label{sec:lowquad}

\begin{table}[htbp]
   \centering
   \begin{tabular}{@{} lccc @{}} 
   \hline
Map &Quadrupole&Probability & Expected\\
&power&& theoretical \\
&($\mu K^2$)&($\%$)&value ($\mu K^2$)\\
\hline
1)\\
W7 Best Fit&       210.3&       3.0\\
W9 Best Fit&       157.7&       1.6\\
\hline
2)\\
TOH W1&       203.2&       2.8\\
ILC W3&       250.5&       4.4\\
ILC W5&        246.9&       4.3\\
ILC W7&       245.4&       4.2\\
ILC W9&       248.2&       4.3\\
TOH W1 (inp)&       154.5&       1.5&1161.3\\
ILC W3 (inp)&       102.1&      0.58\\
ILC W5 (inp)&       117.0&      0.80\\
Dela W5 (inp)&       112.7&      0.73\\
ILC W7 (inp)&       115.4&      0.78\\
ILC W9 (inp)&       124.7&      0.93\\
\hline
3) ISW subtracted \\(measured amplitude\\
2MASS + NVSS)\\
TOH W1 - ISW&      182.5&       4.9\\
ILC W3 - ISW&       189.1&       5.3\\
ILC W5 - ISW&       194.0&       5.6\\
ILC W7 - ISW&       189.6&       5.4\\
ILC W9 - ISW&       194.8&       5.7\\
TOH W1 (inp) - ISW&       129.8&       2.4&800.3\\
ILC W3 (inp) - ISW&       60.7&      0.41\\
ILC W5 (inp) - ISW&       77.6&      0.73\\
Dela W5 (inp) - ISW&       71.5&      0.60\\
ILC W7 (inp) - ISW&       71.9&      0.61\\
ILC W9 (inp) - ISW&       81.7&      0.83\\

\hline

4)\\
W3-ISW\\
2MASS only &600.7&19.8 &1251.8\\
FP10\\

\hline 
   \end{tabular}
   \caption {Quadrupole ($\ell=2$) power and corresponding probability of the quadrupole power being so low. 1): For the best fit of WMAP 7th year and 9th year data. 2): For 11 different CMB maps. 3): After subtraction of the reconstructed ISW signal due to 2MASS and NVSS galaxies. 4): Results from FP10 \citep[][bottom row]{Francis:2010iswanomalies}. Probabilities in this work are calculated using the expected theoretical value given in the last column taken from WMAP 9 best-fit results, whereas FP10 use best-fit results from WMAP 3.} 
   \label{tab:quad}
\end{table}

The low power of the quadrupole was first reported for COBE data \citep{Bennett:1992,Bond:1998,Hinshaw:1996} and subsequently observed in WMAP data \citep{Spergel:2003cb}. Depending on how the significance of the low quadrupole is measured, recent WMAP papers now argue that there is no anomaly in the amplitude of the quadrupole \citep{WmapAnomalies,ilc:w9}. However, we are still interested in studying how the amplitude of the quadrupole behaves after sparse inpainting and after subtraction of the reconstructed ISW field.

The values of the quadrupole before subtraction of the ISW field are reported in the top part of Table \ref{tab:quad} [labelled `1)' and `2)'] before and after sparse inpainting. The third column of Table \ref{tab:quad} gives the probability \footnote{This is done using the idl routine {\tt chisqr\_pdf(quad*df/theory, df), where {\tt quad} is the observed value of the quadrupole, {\tt theory} the theoretical value given by Equation \ref{eq:bestfitquad}, and {\tt df} the number of degrees of freedom for a quadrupole, i.e. 5.}} for a $\chi^2$ random variable with 5 degrees of freedom to take a value less than or equal to the WMAP9 \citep{ilc:w9} expected theoretical (TH) value: \begin{equation}C^{\rm TH}_{\rm W9,\ell=2}= 1161.33~\mu K^2.\label{eq:bestfitquad}\end{equation}

The inpainted maps, which were shown not to introduce a bias in the quadrupole reconstruction (see Section \ref{sec:inpainting} and Appendix \ref{app:inpainting}), have lower quadrupole values than the input maps. This means that the application of inpainting for the CMB reconstruction actually increases the significance of the low quadrupole anomaly for all considered CMB maps.

The third part of Table \ref{tab:quad} [labelled `3)'] shows the measured quadrupole power and probability after subtraction of the ISW reconstructed map due to 2MASS and NVSS galaxies.  The last column shows the new expected theoretical value of the CMB quadrupole, given by: \begin{equation} \hat{C}^{\rm TH}_{\ell=2}=C^{\rm TH}_{\rm W9,\ell=2} - C^{\rm TH}_{\rm ISW, 2MASS,\ell=2} -C^{\rm TH}_{\rm ISW, NVSS,\ell=2},\label{eq:primquad}\end{equation}

The values for $C^{\rm TH}_{\rm ISW, 2MASS,\ell=2}$ and $C^{\rm TH}_{\rm ISW, NVSS,\ell=2}$ are given by the values in Table \ref{tab:isw:th} for the combined NVSS + 2MASS case.
These values are compared with those calculated in FP10 [labelled `4)'] of Table \ref{tab:quad}. 

After ISW subtraction, the quadrupole power decreases for all CMB maps (from $102.1-124.7 \mu K^2$ to $60.7-81.7\mu K^2$, omitting TOH, which has a much larger quadrupole value than all the other maps after inpainting), but since the expected theoretical value also decreases there is nearly no change in the significance of the anomalies before or after ISW subtraction.



We note that the WMAP9 team uses a different approach to measuring the significance of the low quadrupole \citep{ilc:w9}. In any case, our study shows that there is no change in the signifiance of the low quadrupole value after ISW subtraction.

\begin{figure*}[htbp]
   \centering
               \includegraphics[width=9cm,angle=0]{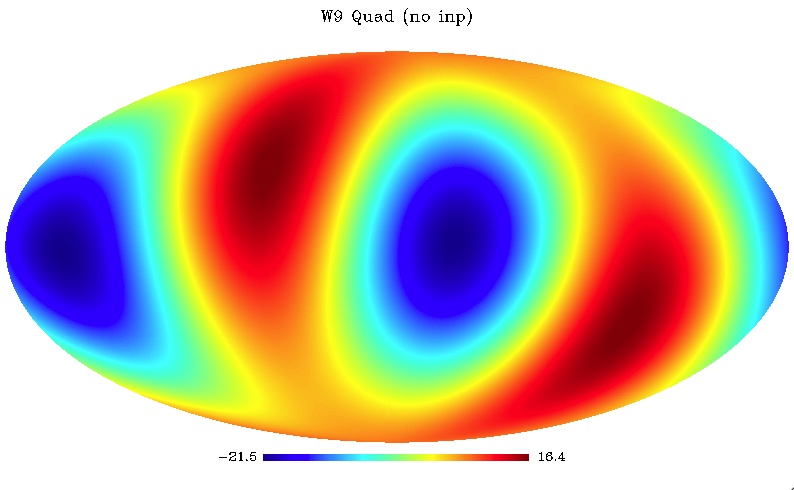}\includegraphics[width=9cm,angle=0]{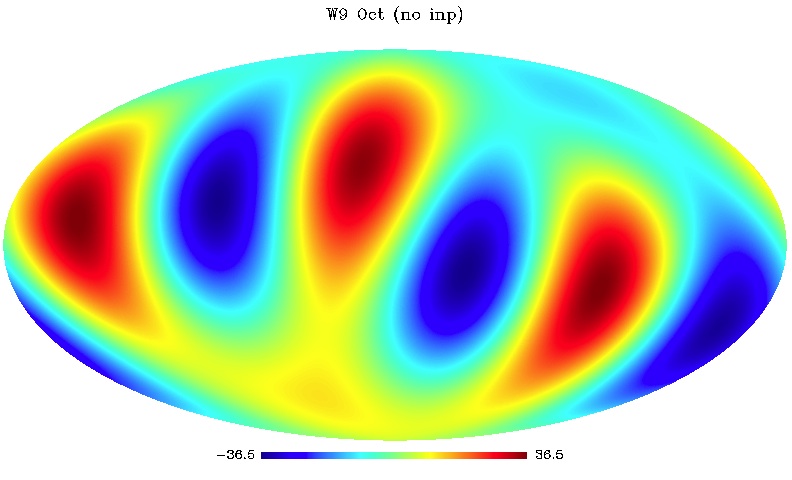}
            \includegraphics[width=9cm,angle=0]{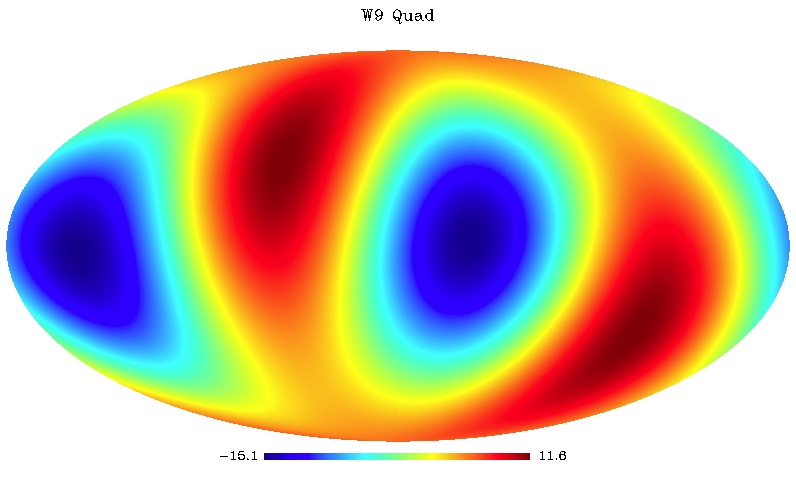}\includegraphics[width=9cm,angle=0]{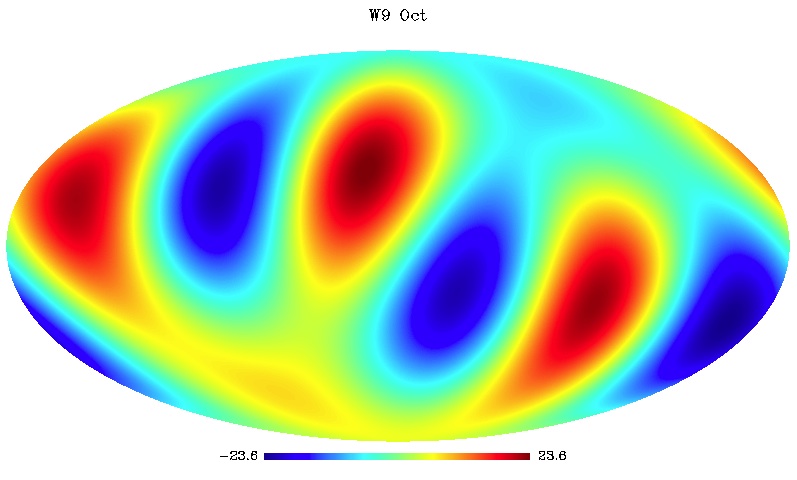}
                        \includegraphics[width=9cm,angle=0]{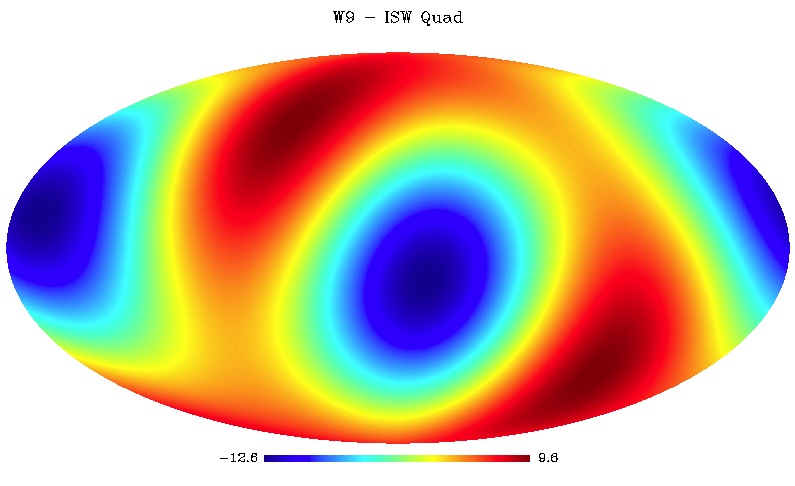}\includegraphics[width=9cm,angle=0]{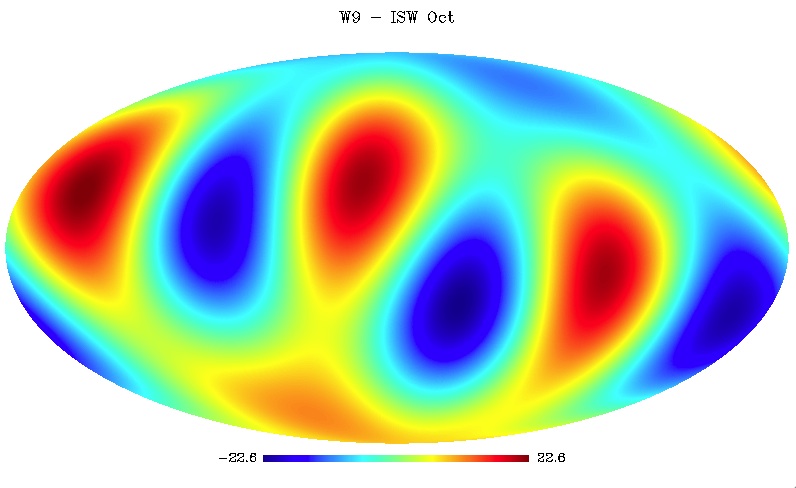}
   \caption{Quadrupole (left) and octopole (right) of WMAP 9 data before inpainting (top), after sparse inpainting (middle), and after subtraction of the reconstructed ISW signal due to 2MASS and NVSS galaxies (bottom).}
   \label{fig:oct_quad}
\end{figure*}
\subsection{Quadrupole/Octopole Alignment}\label{sec:alignment}

\begin{table}[htbp]
   \centering

   \begin{tabular}{@{} lcccccc @{}} 
   \hline
   Map&$\hat{n}_2\cdot \hat{n}_3$&Separation ($^\circ$)&Prob($\%$)\\
 \hline

 TOH            &  0.9856      &   9.7      &   1.4      \\
 W3             &  0.9992      &   2.3      &  0.084 \\
 W5             &  0.9963      &   4.9      &  0.37      \\
 W7             &  0.9966      &   4.7      &  0.34      \\
  W9             &  0.9948      &   5.8      &  0.52      \\
 TOH (inp)      &  0.9808      &   11.2      &   1.9      \\
 W3 (inp)       &  0.9780      &   12.0      &   2.2      \\
 W5 (inp)       &  0.9829      &   10.6      &   1.7      \\
 W5 Dela (inp)  &  0.9514      &   17.9      &   4.9      \\
 W7 (inp)       &  0.9693      &   14.2      &   3.1      \\
 W9 (inp)       &  0.9726      &   13.4      &   2.7      \\
\hline
 After ISW \\subtraction \\(2MASS + NVSS)\\
TOH            - ISW&  0.9113      &   24.3      &   8.9      \\
 W3             - ISW&  0.9535      &   17.0      &   4.6      \\
 W5             - ISW&  0.9381      &   20.3      &   6.2      \\
 W7             - ISW&  0.9414      &   19.7      &   5.9      \\
  W9             - ISW&  0.9260      &   22.2      &   7.4      \\
 TOH (inp)      - ISW&  0.8566      &   31.1      &   14.3      \\
 W3 (inp)       - ISW&  0.9048      &   25.2      &   9.5      \\
 W5 (inp)       - ISW&  0.9095      &   24.6      &   9.1      \\
 W5 Dela (inp)  - ISW&  0.9263      &   22.1      &   7.4      \\
 W7 (inp)       - ISW&  0.9153      &   23.8      &   8.5      \\
 W9 (inp)       - ISW&  0.9088      &   24.7      &   9.1      \\

 \hline

W3 - ISW \\
2MASS only & 0.7548&41.0&24.5\\
FP10\\
\hline

   \end{tabular}
\caption{Scalar product of the preferred axes of the quadrupole and octopole ($\hat{n}_2\cdot \hat{n}_3$), its corresponding separation ($^\circ$), and the probability (\%) of having such a low separation. Note: the theoretically allowed range is $[0^\circ-90^\circ$] since the axes are not vectors. The results are compared with those of FP10 \citep[][bottom row]{Francis:2010iswanomalies}. }
   \label{tab:quadoct}
\end{table}

It was first noted by \cite{OctPlanarity} that not only did the CMB quadrupole and octopole both appear planar (i.e. the dominant mode was $m=\ell$), they also seemed aligned along a similar axis.   For a Gaussian random field there is no reason for the preferred axes of two different multipoles to be correlated.  For each multipole, this preferred axis can be quantified by maximising the quantity 
\begin{equation} q_\ell(\hat{\bf n})=\sum_m m^2 |a_{\ell m}(\hat{\bf{n}})|^2,\end{equation} where $a_{\ell m}(\hat{\bf{n}})$ corresponds to the $a_{\ell m}$ coefficients of the rotated CMB temperature anisotropy map, where $\hat{\bf{n}}$ corresponds to the new z-axis, and is best calculated with at least $nside = 512$. The axes $\hat{\bf n}_2$ and $\hat{\bf n}_3$ correspond to the preferred axes (i.e. where $q_\ell(\hat{\bf n})$ is largest) for $\ell=2$ and $\ell=3$ respectively. We note that this way of measuring the quadrupole/octopole alignment differs from that used by \cite{WmapAnomalies,ilc:w9}, who maximize the quantity: 
\begin{equation} \tilde{q_\ell}(\hat{\bf n}) = |a_{\ell,\ell}|^2+|a_{\ell,-\ell}|^2.\end{equation}

\cite{ilc:w9} found that the limiting factor in measuring the quadrupole/octopole alignment was foreground removal and treatment of the Galactic mask. This motivated us to investigate how the measured quadrupole/octopole behaves after sparse inpainting. 

In the top part of Table \ref{tab:quadoct}, we present the dot product of the preferred axes $\hat{\bf{n}}_2, \hat{\bf{n}}_3$ for $\ell=2,3$ for each CMB map before and after sparse inpainting and before and after ISW subtraction, as well as their corresponding separation (note the maximum separation is $90^\circ$ since the axes are not vectors). We also note the probability of having such a separation in a Gaussian random field, which we estimate using the fact that the dot product of the two vectors has a uniform distribution \citep{OctPlanarity}.  Finally, we compare this with the effect found by FP10 using their 2MASS-ISW reconstruction subtracted from the ILC W3 map (last line of Table \ref{tab:quadoct}).

For non-inpainted maps, the separation varies from 2.3-9.7$^\circ$, i.e. probabilities ranging from 0.08\% to 1.4\%. The same CMB maps, when inpainted, are less anomalous with separations ranging from 10.6-17.9$^\circ$, corresponding to probabilities of 1.7-4.9\%

When the ISW field due to 2MASS and NVSS galaxies is removed, FP10 noted that the alignment was significantly reduced (see the last line in Table \ref{tab:quadoct} or their Table 2 for more details).  Using our reconstruction, we also find the alignment is reduced, though the significance is not as large as in FP10. 
 We note that we are removing the ISW from both 2MASS and NVSS galaxies, whereas FP10 only removed the ISW field due to 2MASS galaxies. Therefore, the general trend is that removing the ISW field reduces the anomalous alignment of the quadrupole and octopole. 

Figure \ref{fig:oct_quad} shows the quadrupole (left) and octopole (right) for the specific case of WMAP9 data, before inpainting (top), after inpainting (middle) and after ISW subtraction (bottom). Inpainting reduces the power in both the WMAP9 quadrupole and octopole, as we had seen in Figure \ref{fig:cmb_quad_oct}. Before inpainting, the preferred axes are in the direction of $(l,b)=(-124.1,57.1)$ and $(-122.3,62.9)$ for $\ell=2,3$ respectively. After inpainting, these become $(l,b)=(-134.5,54.8)$ and $(-110.8, 57.7)$. After inpainting \emph{and} subtraction of the ISW signal due to 2MASS and NVSS galaxies, the preferred axes become $(l,b)=(-90.1,40.8)$ and $(-97.2,65.1)$ for $\ell=2,3$ respectively. In other words, the quadrupole axis is the one which changes the most and which causes the anomaly to decrease since this anomaly measures the correlation between the preferred direction of both the quadrupole and the octopole. We find this trend is true for all maps considered, i.e. that it is the change to the quadrupole shape which is the main cause of the anomaly decrease, whether after sparse inpainting only or after ISW subtraction.

\subsection{Planarity of the Octopole}\label{sec:planarity}
The third anomaly we investigate is the reported planarity of the CMB octopole, i.e. the fact that the phase $m=\ell$ is preferred, which was first noted by \cite{OctPlanarity}. To quantify if the octopole is planar, \cite{OctPlanarity} suggested measuring the quantity: \begin{eqnarray}%
t &=& \max_{\hat{\bf n}} \frac{|a_{3-3}(\hat{\bf n})|^2+|a_{33}(\hat{\bf n})|^2}{\sum_{m=-3}^{m=3}|a_{3m}(\hat{\bf n})|^2}.\end{eqnarray}
This quantity represents the ratio of the octopole power which is contained in the mode $m=\ell=3$, i.e. it is a test of planarity.  For a Gaussian random field, the distribution of power amongst modes should be random, and so there is no reason that the value of $t$ should be close to 1.

In Table \ref{tab:tstat:2mass}, we report values of the `t' statistic for CMB maps, before and after inpainting, and before and after ISW subtraction. The `t' value is calculated using $nside=128$, and its corresponding probability is estimated using 1000 simulated Gaussian random fields with the same power as the map considered.

Before inpainting, the probability of having such planar octopoles ranges from 9.60-14.3\%. \cite{OctPlanarity} reported $t=0.94$ and a probability around $7\%$ using the TOH map, so we first note that later WMAP data are less anomalous than the first year. After inpainting, octopole planarity is reduced even further, with probabilities ranging from $13.6-26.1\%$. After ISW subtraction, FP10 had found that the octopole became even more planar. We find the opposite, i.e. that after ISW subtraction, there is no evidence for any octopole planarity. However, we keep in mind that even before ISW subtraction the significance of the octopole planarity was not particularly significant.

\begin{table}[htbp]
   \centering
   \begin{tabular}{@{} lccc @{}} 
   \hline
   Map & `t' value &Probability (\%)\\
   \hline
 TOH            &   0.9415     &   9.60     &\\
 W3             &   0.9171     &   14.3     &\\
 W5             &   0.9190     &   13.8     &\\
 W7             &   0.9261     &   12.7     &\\
  W9         &   0.9329     &   11.8     &\\
 TOH W1 (inp)   &   0.8917     &   23.4     &\\
 W3 (inp)       &   0.8820     &   26.1     &\\
 W5 (inp)       &   0.8886     &   24.4     &\\
 W5 Dela (inp)  &   0.8919     &   23.4     &\\
 W7 (inp)       &   0.9149     &   15.1     &\\
 W9 (inp)       &   0.9205     &   13.6     &\\
\hline 
TOH            -ISW&   0.8975     &   21.8     &\\
 W3             -ISW&   0.8769     &   27.5     &\\
 W5             -ISW&   0.8776     &   27.4     &\\
 W7             -ISW&   0.8712     &   28.8     &\\
W9         -ISW&   0.8764     &   27.6     &\\
 TOH W1 (inp)   -ISW&   0.8086     &   46.0     &\\
 W3 (inp)       -ISW&   0.7928     &   50.8     &\\
 W5 (inp)       -ISW&   0.7998     &   48.9     &\\
 W5 Dela (inp)  -ISW&   0.80167     &   48.0     &\\
 W7 (inp)       -ISW&   0.8169     &   44.1     &\\
 W9 (inp)       -ISW&   0.8223     &   42.7     &\\

\hline

 \hline
 W3-ISW (due to 2MASS)&0.9841&1.6\\
 FP10\\
 \hline
   \end{tabular}
   \caption{The `t' value for the octopole as defined in \cite{OctPlanarity} using \emph{nside=128}, calculated from the observed CMB maps (\emph{top}) and after subtraction of the ISW field due to 2MASS and NVSS galaxies (\emph{middle}). The probability is determined from 1000 Monte-Carlo simulations and compared with results from FP10 \citep[][bottom row]{Francis:2010iswanomalies}. }
   \label{tab:tstat:2mass}
\end{table}

\begin{table}[htbp]
   \centering
   \begin{tabular}{@{} lcc @{}} 
   \hline
Anomaly &After Sparse&After ISW\\
&Inpainting & Subtraction\\
\hline
\\
Low quad&More anomalous & More anomalous\\
\\
Quad/oct &Less anomalous & Less anomalous\\
Alignment&\\
\\
Oct planarity & Less anomalous & Less anomalous\\
\\
\hline

   \end{tabular}
   \caption{Summary of results in this paper. We identify trends from the 11 WMAP data sets described in Table \ref{tab:cmbdata} regarding how sparse inpainting and subtraction of the ISW signal due to 2MASS and NVSS galaxies can affect the three anomalies investigated. The third test for octopole planarity did not return significantly anomalous results in the first place.}
   \label{tab:summary}
\end{table}

\begin{table*}[htbp]
   \centering
   \begin{tabular}{@{} lcc @{}} 
   Product Name&Type & Description\\
   \hline
  kinetic Doppler products:\\
   kDoppler map & Map & kinetic Doppler map for nside=512\\
   kDoppler code & code (IDL) & generates kinetic Doppler map\\
   \hline
   CMB and LSS products:\\
   TOH W1 (inp)&Map&Sparsely inpainting TOH map \\
   ILC W3 (inp)&Map& Sparsely inpainting ILC W3 map\\
   ILC W5 (inp)&Map&Sparsely inpainting ILC W5 map\\
   Dela W5 (inp)&Map&Sparsely inpainting Delabrouille W5 map\\
   ILC W7 (inp) &Map&Sparsely inpainting ILC W7 map\\
   ILC W9 (inp) &Map&Sparsely inpainting ILC W9 map\\
   2MASS (inp) & Map &Sparsely inpainting 2MASS map\\
   NVSS (inp) & Map &Sparsely inpainting NVSS map\\
   {\tt cmb\_lowl\_alm\_inpainting}&code (IDL)& inpaints CMB or LSS maps (compatible with {\tt ISAP}\footnote{\url{http://jstarck.free.fr/isap.html}})\\
   \hline
   2MASS ISW products:\\
   ISW\_2MASS\_TOH& Map &ISW from 2MASS and TOH\\
   ISW\_2MASS\_W1 &Map &ISW from 2MASS and W1\\
   ISW\_2MASS\_W3& Map &ISW from 2MASS and W3\\
   ISW\_2MASS\_W5& Map &ISW from 2MASS and W5\\
   ISW\_2MASS\_W5Dela& Map &ISW from 2MASS and W5 Dela\\
   ISW\_2MASS\_W7& Map &ISW from 2MASS and W7\\
   ISW\_2MASS\_W9& Map &ISW from 2MASS and W9\\
   \hline 
      NVSS ISW products:\\
   ISW\_NVSS\_TOH& Map &ISW from NVSS and TOH\\
   ISW\_NVSS\_W1 &Map &ISW from NVSS and W1\\
   ISW\_NVSS\_W3& Map &ISW from NVSS and W3\\
   ISW\_NVSS\_W5& Map &ISW from NVSS and W5\\
   ISW\_NVSS\_W5Dela& Map &ISW from NVSS and W5 Dela\\
   ISW\_NVSS\_W7& Map &ISW from NVSS and W7\\
   ISW\_NVSS\_W9& Map &ISW from NVSS and W9\\
   \hline 
   Statistics products:\\
   anomalies\_l2l3 & code (F90) & calculates quad/oct alignment and probability (requires {\tt HealPix})\\
   anomalies\_octplan & code (F90) & calculates octopole planarity and probability (requires {\tt HealPix}) \\

   \hline
   
   \end{tabular}
   \caption{List of products made available in this paper in the spirit of reproducible research, available here: \url{http://www.cosmostat.org/anomaliesCMB.html}.}
   \label{tab:reproducible}
\end{table*}

\section{Discussion}\label{sec:discussion}
 
Although there is currently a debate over the significance of the claimed large-scale anomalies in the CMB \citep[see e.g.,][]{WmapAnomalies}, their existence is not totally dismissed. In parallel to the debate on the statistical significance of the anomalies, recent work has also focussed on the impact of the reconstruction method to account for masked regions of the sky \citep[e.g.,][]{Bernui:2006,Slosar:2004s,Kim:2012}. Moreover, in some cases the Galactic mask is a limiting factor in the study of anomalies \citep{ilc:w9}. Studies have also focussed on low-redshift cosmology and astrophysics as potential sources of contamination \citep{Rudnick:2007,Hiranya:ksz,Francis:2010iswanomalies,CMBSN:2012}.  Since statistical isotropy is predicted for the early Universe, analyses should focus on the primordial CMB, i.e. one from which secondary low-redshift cosmological signals have been removed. This should be done \emph{whether or not} one believes the significance of the reported anomalies in the CMB.

In this paper, we focus simultaneously on both the reconstruction method and the ISW effect as a means to estimate the primoridal CMB and test it for anomalies.  We focus on three previously reported large-scale anomalies, namely: the low quadrupole power, the quadrupole/octopole alignment, and the octopole planarity. 

This work updates that of \cite{Francis:2010iswanomalies} and \cite{Kim:2012} in the following ways: \begin{enumerate}
\item \emph{Tomography}: We reconstruct the ISW signal due to both 2MASS \citep{Jarrett:2000me} and NVSS \citep{NVSS} data (see Section \ref{sec:data})
\item \emph{Trends}: We look for trends by considering 11 different WMAP data sets (see Section \ref{sec:data}), including 6 data sets for which we reconstructed missing data using the sparse inpainting technique mentioned below

\item \emph{Sparse inpainting}: Missing data in CMB and LSS maps are reconstructed using sparse inpainting (see Section \ref{sec:inpainting}, A08-SMF10), which does not that assume the data are Gaussian or statistically isotropic. 
\item \emph{Tests for biases}: We show in Section \ref{sec:inpainting} and Appendix \ref{app:inpainting} that the sparse inpainting method used is a bias-free reconstruction, which does not introduce a spurious statistical anisotropies nor a spurious ISW signal 
\end{enumerate}

In this work we first investigate the impact of our bias-free sparse inpainting on various CMB maps and on the claimed CMB anomalies. We then subtract the ISW signal reconstructed in Section \ref{sec:inpainting} and test again for impact on the three studied anomalies. Our main conclusions are made by identifying trends amongst the 11 CMB maps considered. 

Our conclusions are summarised in Table \ref{tab:summary}, namely: 
\begin{itemize}
\item The low quadrupole becomes more anomalous after sparse inpainting of CMB maps, and remains so after additional subtraction of the ISW signal, contrarily to what \cite{Francis:2010iswanomalies} (who used a different reconstruction method) found.
\item The quadrupole/octopole alignment becomes less anomalous after sparse inpainting of the CMB maps, contrary to what was reported by \cite{Kim:2012}, who used a Gaussian constrained inpainting method. 
\item The quadrupole/octopole alignment anomalies are reduced in significance after subtraction of the ISW signal, similarly to what was reported by \cite{Francis:2010iswanomalies}.
\item We note that the reduced significance of the quadrupole/octopole alignment (both after sparse inpainting and ISW subtraction) is mainly due to changes in the quadrupole shape, not the octopole, suggesting that the main source of both anomalies could be the quadrupole. 
\item We find that after inpainting and ISW subtraction, the octopole planarity becomes less anomalous, contrarily to the report of \cite{Francis:2010iswanomalies} who found the octopole planarity  had become more anomalous after ISW subtraction. We also note that the octopole planarity did not seem significant in any CMB map subsequent to the first year WMAP data.
\end{itemize}

We therefore conclude that after application of sparse inpainting and subtraction of the ISW signal due to the low-redshift Universe out to $z\sim 1$, as estimated from the 2MASS and NVSS surveys, that the quadrupole/octopole alignment and the octopole planarity appear less significant. In addition, it seems that both the low quadrupole and quadrupole/octopole alignment could be in fact due to the quadrupole only. Other hypotheses remain possible (e.g. exotic physics).

In the spirit of participating in reproducible research, we make public all codes and resulting products which constitute the main results of this paper public. In Table \ref{tab:reproducible} we list the products which are made freely available as a result of this paper and which are available here: \url{http://www.cosmostat.org/anomaliesCMB.html}.

\begin{acknowledgements}
The authors are grateful to John Peacock, Caroline Francis, Shirley Ho, Olivier Dor\'e, Arnaud Woiselle, the Euclid CMB cross-correlations working group and the anonymous referee for useful discussions. AR thanks Fran\c cois Lanusse, Florent Sureau, and Yves Revaz for computational help. We also thank Jacques Delabrouille for providing the wavelets ILC 5yr map from the WMAP 5yr data.  We used iCosmo \footnote{\url{http://www.icosmo.org}} software \citep{Refregier:2011}, Healpix software \citep{healpix:2002,Gorski:2004by},
  ISAP\footnote{\url{http://jstarck.free.fr/isap.html}} software, the 2 Micron All-Sky Survey catalogue (2MASS)
  \footnote{\url{http://www.ipac.caltech.edu/2mass/}}, \emph{WMAP} data \footnote{\url{http://map.gsfc.nasa.gov}}, the NRAO VLA Sky Survey  \citep[NVSS,][]{NVSS} \footnote{\url{http://heasarc.gsfc.nasa.gov/W3Browse/all/nvss.html}},
  and the Galaxy extinction maps of \cite{Schlegel:1997yv}. This work was supported by the European Research Council grant SparseAstro (ERC-228261) and by the Swiss National Science Foundation
(SNSF).
\end{acknowledgements}

\appendix

\section{Testing statistical isotropy biases for sparse inpainting}\label{app:inpainting}
To justify that sparse inpainting is indeed an appropriate reconstruction method in the context of probing anomalies in the primordial CMB, we test whether we can indeed consider sparse inpainting as a bias-free reconstruction. Tto do this we test whether the three statistics studied in Section \ref{sec:probing} are unchanged after the application of sparse inpainting to CMB, LSS, and ISW simulated maps. 

Our aim here is to test whether statistically isotropic maps conserve this quality after sparse inpainting. It would also be interesting to test whether maps which are intrinsically anomalous are still anomalous after sparse inpainting. However, testing the latter requires a knowledge of the model creating the anomalies (exotic physics, foregrounds, etc\ldots) in order to test these realistically.  In this appendix, we focus only on testing whether sparse inpainting creates spurious statistically \emph{anisotropic} signatures in data where the true distribution is known to be statistically \emph{isotropic}. 

We sparsely inpaint all CMB and LSS simulations using the public {\tt ISAP} software \footnotemark[8] of A08-SMF10 as described in section \ref{sec:inpainting}.

\subsection{Impact of Sparse Inpainting on CMB data}
We start by considering the impact of sparse inpainting on CMB maps. We consider 1000 Gaussian random field realisations of a CMB map with an input power spectrum corresponding to the ``vanilla'' cosmology considered in Section \ref{sec:probing}. We sparsely inpaint each simulation after applying the WMAP 7th year temperature analysis mask, as described in Section \ref{sec:data} and \ref{sec:inpainting}.

We test the mean values for each statistic (low quadrupole, quadrupole/octopole alignment, and octopole planarity) before and after sparse inpainting. The results are reported in Table \ref{tab:app:cmb2} and show that the sparse inpainting method introduces negligible biases in all three tests, compared the intrinsic cosmic variance.  The test of quadrupole/octopole alignment can be somewhat altered due to sparse inpainting (see the standard deviation on the bias).

Since with current data we observe a low quadrupole, we test whether sparse inpainting is also robust in cases where the quadrupole is low.  We perform a new set of 400 Gaussian random field simulations, where the input power spectrum is WMAP 7 data. Table \ref{tab:app:cmb1} shows the results for the three statistical isotropy tests performed on this set of simulations before and after application of sparse inpainting. Again, we see that biases are very small compared to cosmic variance.

\subsection{Impact of Sparse Inpainting on LSS data}
For the reconstruction of the ISW signal, we must first reconstruct full-sky LSS maps (Section \ref{sec:inpainting}). Thus we also want to test whether the statistics of the LSS maps are altered by sparse inpainting. To do this, we specifically test how Gaussian random field simulations of 2MASS and NVSS-like surveys are affected by sparse inpainting using the same masks as used in Section \ref{sec:data}. The simulated maps are for the galaxy overdensity (i.e. including the galaxy bias). Results are shown in Tables \ref{tab:app:2mass} (2MASS) and \ref{tab:app:nvss} (NVSS).  Again, biases are negligible compared to cosmic variance for all three statistics considered.

\subsection{Impact of Sparse Inpainting on the ISW signal}
Finally, we want to test whether sparse inpainting can alter the reconstructed ISW signal, either by modifying tests of statistical isotropy or by introducing a spurious ISW signal. We consider 1000 Gaussian random realisations of NVSS galaxies and 1000 \emph{uncorrelated} WMAP 7 realisations. The goal is to see if inpainting of each map with its respective masks can introduce a spurious ISW signal, when none should be measured. The results are shown in the first line of Table \ref{tab:app:nvssisw}, which gives the value of the cross-correlation quadrupole ($C_{gT,\ell=2}$). Since the maps are uncorrelated by construction, we expect a median cross-correlation equal to zero, which is what we find both before and after sparse inpainting. This indicates that our method does not introduce a spurious cross-correlation signal. 

We also test the quadrupole/octopole alignment and octopole planarity of the reconstructed ISW signal from NVSS galaxies (where the ISW map is constructed from NVSS and CMB simulations with different masks). We find that the tests are unbiased after sparse inpainting.

\begin{table}[htbp]
   \begin{tabular}{@{} lccc @{}} 
   \hline
 &ILC Th & ILC Th& \\
 &simulations & inpainted \\
 &&simulations\\
 &$a$&$b$&$a-b$\\
 \hline
 Expected&\\
 Quadrupole& $      1269.0\pm  806.0$&$  1229.9\pm       855.1$&$       39.1\pm       334.3$\\

$[\mu K^2]$ \\
 \\
Expected  \\

 $\left< {\bf \hat{n}_2}\cdot {\bf \hat{n}_3}\right>$&$0.489\pm0.302$&$0.503\pm0.296$&$-0.014\pm0.181$\\

 \\
Expected & \\
$\left<t\right>$ value&$0.78\pm0.117$&$0.78\pm0.119$&$-0.0012\pm0.061$\\
\\
 \hline
   \end{tabular}
   \caption{Testing whether sparse inpainting (A08-SMF10) affects the tests of statistical isotropy used in Section \ref{sec:probing} for 1000 simulated Gaussian random fields with input power spectrum taken as the best-fit theoretical power spectrum for WMAP7 (i.e. \emph{with no} low quadrupole). }
   \label{tab:app:cmb2}
\end{table}

\begin{table}[htbp]
   \begin{tabular}{@{} lccc @{}} 
   \hline
 &ILC W7 & ILC W7& \\
 &simulations & inpainted \\
 &&simulations\\
 &$a$&$b$&$a-b$\\
 \hline
Expected&\\
Quadrupole&       $252.1\pm        157.0$& $      244.9\pm       175.8$&    $  7.2\pm       85.0$\\

$[\mu K^2]$ \\
 \\
Expected  \\
 $\left< {\bf \hat{n}_2}\cdot {\bf \hat{n}_3}\right>$&$0.510\pm 0.292$&$0.526\pm0.292$&$-0.0156 \pm  0.207$\\ 
 \\
 \\
 
Expected & \\

$\left<t\right>$ value& $0.79  \pm  0.12$&$0.80 \pm  0.12$&$ 0.0037 \pm 0.065$\\ 
\\
 \hline
   \end{tabular}
   \caption{Testing whether sparse inpainting (A08-SMF10) affects the three tests of statistical isotropy used in Section \ref{sec:probing}  for 400 simulated Gaussian random fields with input power spectrum taken as WMAP7-like data (i.e. \emph{with} low quadrupole).  }
   \label{tab:app:cmb1}
\end{table}

\begin{table}[htbp]
   \begin{tabular}{@{} lccc @{}} 
   \hline
 &2MASS gal & 2MASS gal& \\
 &simulations & inpainted \\
 &&simulations\\
 &$a$&$b$&$a-b$\\
 \hline
Expected&\\

Quadrupole &$2.96\pm1.93$  &$3.00 \pm 2.49$&$-0.041\pm 1.44$\\
 $[\times 10^{-3}$]\\
 \\
 
Expected  \\
 $\left< {\bf \hat{n}_2}\cdot {\bf \hat{n}_3}\right>$&$0.546\pm0.290$&$0.536\pm0.303$&$0.011\pm0.392$\\

 \\
Expected &\\
$\left<t\right>$ value&$0.782\pm0.120$&$0.801\pm0.115$&$-0.019\pm0.145$\\

 \hline
   \end{tabular}
   \caption{Testing whether sparse inpainting (A08-SMF10) affects the tests of statistical isotropy used in Section \ref{sec:probing} for 1000 simulated Gaussian random fields with input power spectrum taken as the theoretical galaxy power spectrum for 2MASS galaxies.}
   \label{tab:app:2mass}
\end{table}

\begin{table}[htbp]
   \begin{tabular}{@{} lccc @{}} 
   \hline
 &NVSS gal & NVSS gal& \\
 &simulations & inpainted \\
 &&simulations\\
 &$a$&$b$&$a-b$\\
 \hline
Expected &  \\

Quadrupole&$2.55\pm1.58$&$2.96\pm2.56$&$-0.405\pm 1.97$\\
$[\times 10^{-6}]$\\
 \\
Expected  \\
 $\left< {\bf \hat{n}_2}\cdot {\bf \hat{n}_3}\right>$&$0.530\pm0.295$&$0.512\pm0.291$&$0.0182\pm0.395$\\
 \\

Expected & \\
$\left<t\right>$ value& $0.783\pm0.114$&$0.796\pm0.115$&$-0.0133\pm0.153$\\
\\
 \hline
   \end{tabular}
   \caption{Testing whether sparse inpainting (A08-SMF10) affects the tests of statistical isotropy used in Section \ref{sec:probing} for 1000 simulated Gaussian random fields with input power spectrum taken as the theoretical galaxy power spectrum for NVSS galaxies.}
   \label{tab:app:nvss}
\end{table}

\begin{table}[htbp]
   \begin{tabular}{@{} lccc @{}} 
   \hline
 &NVSS ISW &NVSS ISW& $a-b$\\
 &simulations & inpainted \\
 &&simulations\\
 &a&b&\\
 \hline

$C_{gT, \ell=2}$&$-4.10^{-4}\pm0.025$&$1.10^{-4}\pm0.026$&$ -5.10^{-4}\pm0.023$\\
\\
Expected  \\
 $\left< {\bf \hat{n}_2}\cdot {\bf \hat{n}_3}\right>$&$0.546\pm0.290$&$0.536\pm0.303$&$0.0109\pm0.392$\\
 \\

Expected & \\
$\left<t\right>$ value& $0.783\pm0.114$&$0.791\pm0.114$&$-0.007\pm0.155$\\
\\

 \hline
   \end{tabular}
   \caption{Expected values of all anomaly tests presented in this paper for 1000 ISW maps due to NVSS galaxies. The ISW maps are calculated from both CMB and NVSS Gaussian random field simulations using Equation \ref{eq:alm_isw}. The first column shows results for full sky simulations, while the second column shows results for inpainting CMB and NVSS simulations with respective masks as described in Section \ref{sec:inpainting}.}
   \label{tab:app:nvssisw}
\end{table}

\bibliographystyle{aa}
\bibliography{./article.bib}

\end{document}